\documentclass[letterpaper,twocolumn,aps,superscriptaddress,amsmath,amssymb,floatfix]{revtex4-1}
\usepackage{mathptmx}
\usepackage[utf8]{inputenc}
\usepackage{color}
\usepackage{amsmath}
\usepackage{amssymb}
\usepackage{graphicx}
\usepackage{esint}
\usepackage[unicode=true,
 bookmarks=true,bookmarksnumbered=false,bookmarksopen=false,
 breaklinks=false,pdfborder={0 0 1},backref=false,colorlinks=true]
 {hyperref}
\hypersetup{
 linkcolor=magenta,urlcolor=blue,citecolor=blue,pdfstartview={FitH},hyperfootnotes=false}

\makeatletter

\usepackage{textcomp}
\usepackage{epstopdf}

\pdfpageheight\paperheight
\pdfpagewidth\paperwidth

\@ifundefined{textcolor}{}{%
 \definecolor{BLACK}{gray}{0}
 \definecolor{WHITE}{gray}{1}
 \definecolor{RED}{rgb}{1,0,0}
 \definecolor{GREEN}{rgb}{0,1,0}
 \definecolor{BLUE}{rgb}{0,0,1}
 \definecolor{CYAN}{cmyk}{1,0,0,0}
 \definecolor{MAGENTA}{cmyk}{0,1,0,0}
 \definecolor{YELLOW}{cmyk}{0,0,1,0}
}

\usepackage{xcolor}\usepackage{soul}
\definecolor{blue}{rgb}{0,0,1}
\definecolor{red}{rgb}{1,0,0}
\definecolor{green}{rgb}{0,1,0}

\usepackage{soul}

\makeatother

\begin{document}
\title{Compact and high-resolution spectrometer via Brillouin integrated circuits}

\author{Jia-Qi~Wang}
\thanks{These three authors contributed equally to this work.}
\affiliation{CAS Key Laboratory of Quantum Information, University of Science and Technology of China, Hefei 230026, P. R. China.}
\affiliation{Anhui Province Key Laboratory of Quantum Network, University of Science and Technology of China, Hefei 230026, P. R. China}

\author{Yuan-Hao~Yang}
\thanks{These three authors contributed equally to this work.}
\affiliation{CAS Key Laboratory of Quantum Information, University of Science and Technology of China, Hefei 230026, P. R. China.}
\affiliation{Anhui Province Key Laboratory of Quantum Network, University of Science and Technology of China, Hefei 230026, P. R. China}

\author{Zheng-Xu~Zhu}
\thanks{These three authors contributed equally to this work.}
\affiliation{CAS Key Laboratory of Quantum Information, University of Science and Technology of China, Hefei 230026, P. R. China.}
\affiliation{Anhui Province Key Laboratory of Quantum Network, University of Science and Technology of China, Hefei 230026, P. R. China}

\author{Juan-Juan Lu}
\affiliation{School of Information Science and Technology, ShanghaiTech University, Shanghai 201210, China.}
\author{Ming~Li}
\affiliation{CAS Key Laboratory of Quantum Information, University of Science and Technology of China, Hefei 230026, P. R. China.}
\affiliation{Anhui Province Key Laboratory of Quantum Network, University of Science and Technology of China, Hefei 230026, P. R. China}

\author{Xiaoxuan~Pan}
\affiliation{Center for Quantum Information, Institute for Interdisciplinary Information
Sciences, Tsinghua University, Beijing 100084, China}

\author{Chuanlong~Ma}
\affiliation{Center for Quantum Information, Institute for Interdisciplinary Information
Sciences, Tsinghua University, Beijing 100084, China}

\author{Lintao~Xiao}
\affiliation{Center for Quantum Information, Institute for Interdisciplinary Information
Sciences, Tsinghua University, Beijing 100084, China}

\author{Bo~Zhang}
\affiliation{Center for Quantum Information, Institute for Interdisciplinary Information
Sciences, Tsinghua University, Beijing 100084, China}

\author{Weiting~Wang}
\affiliation{Center for Quantum Information, Institute for Interdisciplinary Information
Sciences, Tsinghua University, Beijing 100084, China}

\author{Chun-Hua~Dong}
\affiliation{CAS Key Laboratory of Quantum Information, University of Science and Technology of China, Hefei 230026, P. R. China.}
\affiliation{Anhui Province Key Laboratory of Quantum Network, University of Science and Technology of China, Hefei 230026, P. R. China}

\author{Xin-Biao~Xu}
\email{xbxuphys@ustc.edu.cn}
\affiliation{CAS Key Laboratory of Quantum Information, University of Science and Technology of China, Hefei 230026, P. R. China.}
\affiliation{Anhui Province Key Laboratory of Quantum Network, University of Science and Technology of China, Hefei 230026, P. R. China}

\author{Guang-Can~Guo}
\affiliation{CAS Key Laboratory of Quantum Information, University of Science and Technology of China, Hefei 230026, P. R. China.}
\affiliation{Anhui Province Key Laboratory of Quantum Network, University of Science and Technology of China, Hefei 230026, P. R. China}

\author{Luyan~Sun}
\email{luyansun@tsinghua.edu.cn}
\affiliation{Center for Quantum Information, Institute for Interdisciplinary Information
Sciences, Tsinghua University, Beijing 100084, China}

\author{Chang-Ling~Zou}
\email{clzou321@ustc.edu.cn}
\affiliation{CAS Key Laboratory of Quantum Information, University of Science and Technology of China, Hefei 230026, P. R. China.}
\affiliation{Anhui Province Key Laboratory of Quantum Network, University of Science and Technology of China, Hefei 230026, P. R. China}
\affiliation{CAS Center For Excellence in Quantum Information and Quantum Physics,
University of Science and Technology of China, Hefei, Anhui 230026,
P. R. China.}

\begin{abstract}
\textbf{Optical spectrometers are indispensable tools across various fields, from chemical and biological sensing to astronomical observations and quantum technologies. However, the integration of spectrometers onto photonic chips has been hindered by the low spectral resolution or large device footprint with complex multiple channel operations. Here, we introduce a novel chip-integrated spectrometer by leveraging the acoustically-stimulated Brillouin scattering in a hybrid photonic-phononic chip. The Brillouin interaction provides a dynamic reflection grating with a high reflectivity up to 50\% and a fast switching time on the microsecond scale, achieving an unprecedented spectral resolution of 0.56\,nm over a 110\,nm bandwidth using just a single 1\,mm-long straight waveguide. This remarkable performance approaches the fundamental limit of resolution for a given device size, validating the potential of the hybrid photonic-phononic device for efficient and dynamically-reconfigurable spectral analysis, and thus opens up new avenues for advanced optical signal processing and sensing applications.}
\end{abstract}
\maketitle

\noindent \textbf{\large{}Introduction}{\large\par} 
\noindent Photonic integrated circuits (PICs) have emerged as a powerful platform~\cite{elshaari2020hybrid,wang2020integrated,Shekhar2024,pelucchi2022potential,kippenberg2008cavity,wang2018integrated}, which has the potential to revolutionize the way we generate, process, and manipulate light. By integrating various optical components onto a single chip, PICs enable the realization of sophisticated optical systems that offer unparalleled scalability, low power consumption, high stability, and enhanced light-matter interaction. In recent years, significant progress has been made in the development of PIC technologies, leading to notable achievements in various fields, such as ultra-high data transmission rates in optical communications~\cite{corcoran2020ultra}, ultra-sensitive biosensors~\cite{shen2014silicon}, and quantum photonic devices for secure communication and computation~\cite{eggleton2019brillouin}. To further unlock the capability of PIC in spectroscopic analysis for astronomy measurements~\cite{jovanovic2023,yu2018silicon}, medical diagnostics~\cite{kong2015raman}, environmental monitoring~\cite{ho2005overview}, and chemical analysis~\cite{kneipp1999ultrasensitive,Harper24}, high-performance optical spectrometers are indispensable~\cite{science2021miniaturization,bacon2004miniature,crocombe2018portable}.

\begin{figure*}[!t]
\begin{centering}
\includegraphics[width=1\textwidth]{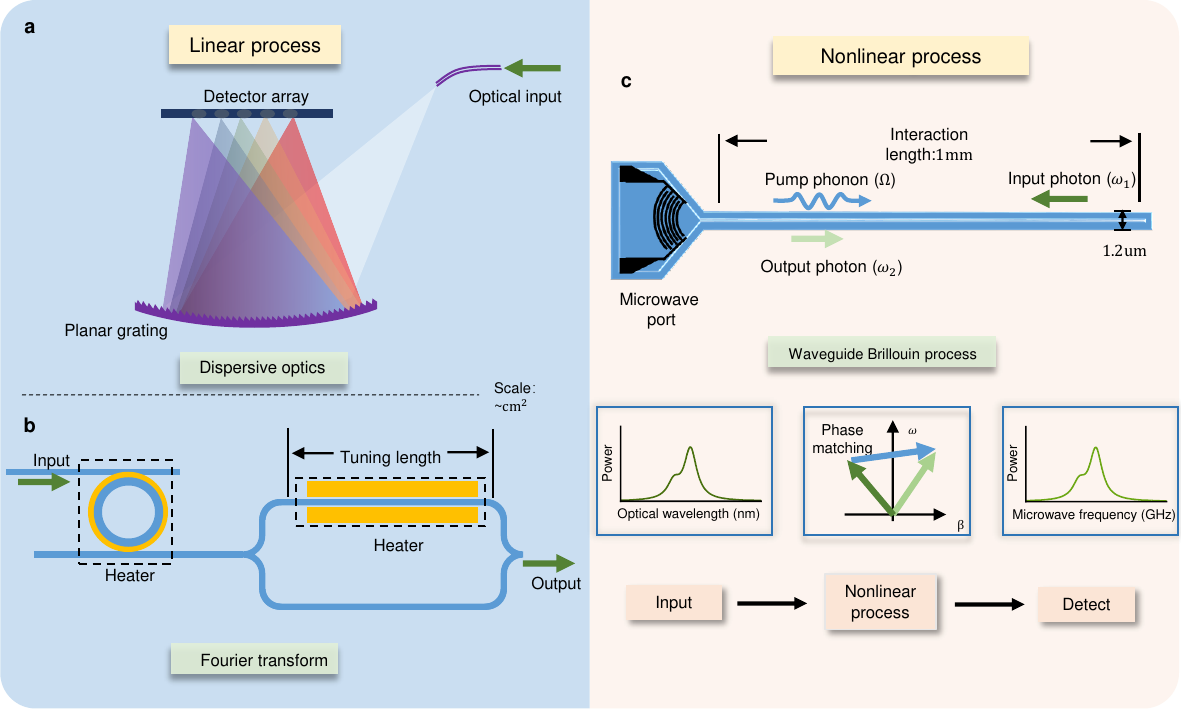}
\par\end{centering}
\caption{\textbf{Schematic of on-chip spectrometer.} \textbf{a}-\textbf{b}, Conventional on-chip spectrometers based on linear processes. \textbf{a}, A static planar grating enables dispersive optics to separate light with different wavelengths. \textbf{b}, Wavelength of the input light can be derived through a Fourier-transform approach by dynamically tuning a microring resonator or a Mach-Zehnder interferometer. \textbf{c}, The proposed spectrometer is based on nonlinear process that employs the phase-matching condition of acoustically-stimulated Brillouin scattering to achieve wavelength-dependent optical reflection. The input laser ($\omega_{1}$) and phonon drive ($\Omega$) propagate in opposite directions, and the reflected light ($\omega_{2}$) satisfies both momentum and energy conservations.}
\label{Fig1}
\end{figure*}

However, the full integration of spectrometers onto photonic chips faces a fundamental trade-off between spectral resolving capability and device footprint~\cite{science2021miniaturization}. Conventional spectrometers mostly rely on static dispersive elements, which could be directly extended to PICs~\cite{calafiore2014holographic,faraji2018compact,cheng2019broadband}. For example, the grating shown in Fig.~\ref{Fig1}a separates light of different colors into distinct channels. While the grating offers a wide detection wavelength range, achieving high spectral resolution necessitates a large device length and a dense array of collection channels. The spectral resolution of a grating-based spectrometer is approximately given by $\Delta_\lambda={\lambda^2}/{n_{\mathrm{eff}}L\left(\mathrm{sin}\theta_1+\mathrm{sin}\theta_2\right)}$, where $L$ is the grating length, $n_{\mathrm{eff}}$ is the effective refractive index of the guided optical mode, $\lambda$ is the light wavelength, and $\theta_{1(2)}$ is the angle between the input (output) optical beam and the grating. An alternative approach is to employ dynamically tunable dispersive devices, such as Mach-Zehnder interferometers~\cite{kita2018high,pohl2020integrated,finco2024monolithic} shown in Fig.~\ref{Fig1}b. These devices offer potential advantages including higher optical throughput, reduced device footprint, and requirement of fewer detection channels. Nonetheless, the resolution of this approach is limited by the device's tuning capability. The spectral resolution of a tunable interferometer-based spectrometer is given by  $\Delta_\lambda={\lambda^2}/{n_{\mathrm{eff}}\Delta L}$ with $\Delta L$ is the tuning path difference of the interferometer. The requirement for a large tuning length $\delta n\Delta L\gg \lambda$, where $\delta n$ is the variation of the material refractive index. This requirement imposes a practical limitation of the spectrometer size $L$, limiting it to the order of centimeters, even with the enhancement provided by microrings~\cite{zheng2019microring}. Consequently, there is a pressing need for an approach that combines the advantages of both static and dynamic dispersive elements, enabling the realization of compact, single-channel on-chip spectrometers.

Here, we introduce a fundamentally different approach to integrated spectroscopy that leverages acoustically-stimulated backward Brillouin scattering, a nonlinear optical interaction between photons and acoustic phonons in a medium (Fig.~\ref{Fig1}c). By harnessing the strong interaction between guided optical and acoustic modes in a carefully engineered lithium niobate (LN) waveguide, we demonstrate a high-resolution, compact, and fully integrated spectrometer with an intrinsic spectral resolution of 0.56\,nm over a wavelength range exceeding 110\,nm, while maintaining a compact footprint of just 1 mm in length. Our approach provides a coherent interface between the optical spectrometer and radio frequency (RF) control, allowing fast spectral sweeping, with a typical sweep duration of $0.1$ seconds and could potentially be even faster. Furthermore, our demonstration of a hybrid photonic-phononic circuit can be extended to incorporate other photonic, electronic, and phononic devices on a single chip, paving the way for fully integrated and multi-functional spectroscopic systems and also for a wide range of novel photonic-phononic hybrid devices.

\begin{figure*}[!t]
\begin{centering}
\textcolor{red}{\includegraphics[width=1\textwidth]{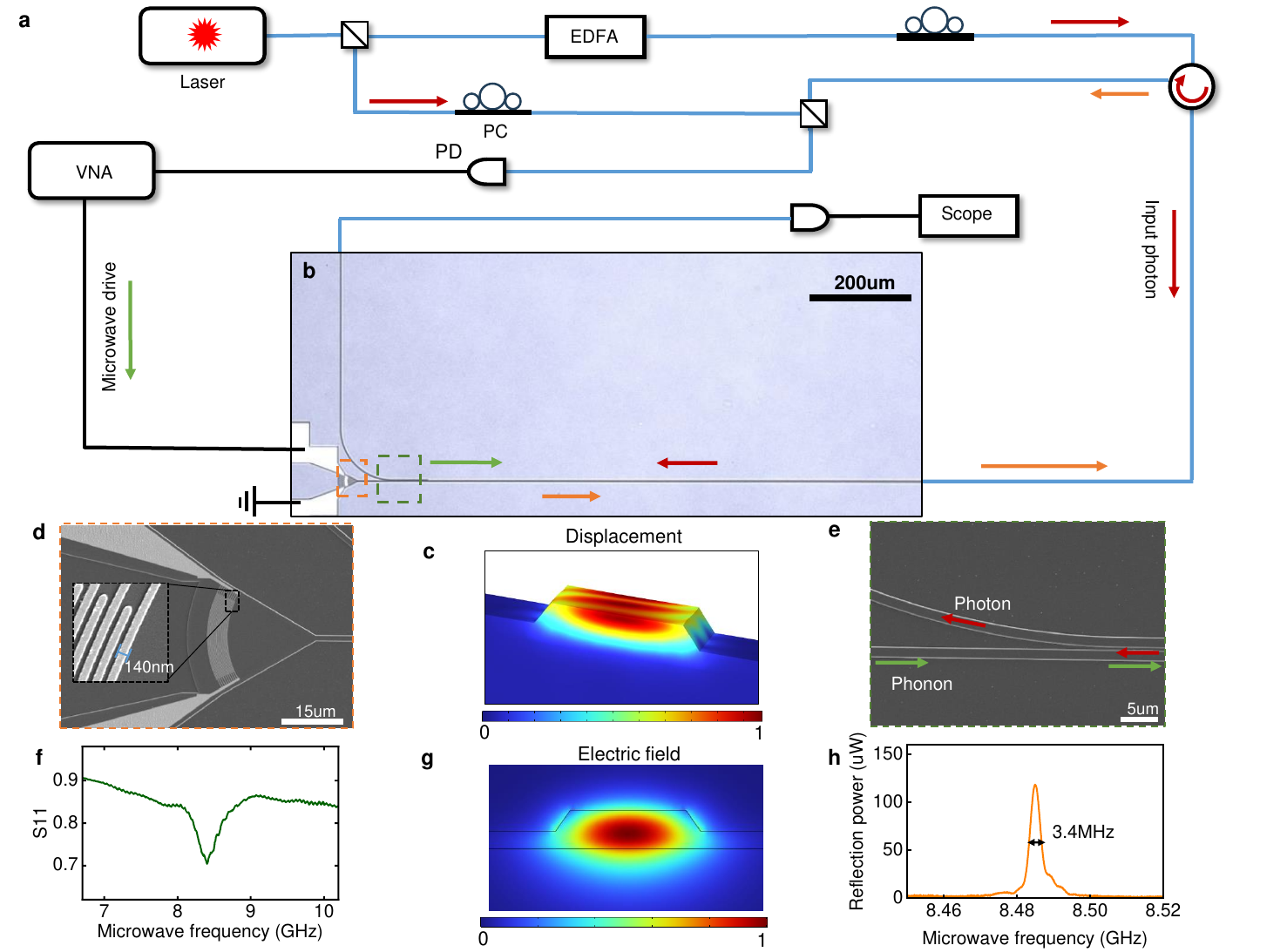}}
\par\end{centering}
\caption{\textbf{Characterization of the on-chip spectrometer.} \textbf{a}-\textbf{b}, Experimental setup for characterizing the spectrometer. The phononic mode in the waveguide is driven by a vector network analyzer (VNA) through an interdigital transducer (IDT, \textbf{d}). The optical input signal is reflected through the Brillouin process when the phase-matching condition is fulfilled. The reflected light is measured by the VNA through a heterodyne detection or directly detected by the photodiode (PD) under high signal-to-noise ratio conditions. The Brillouin interaction region is around $1\,\mathrm{mm}$. EDFA: erbium-doped fiber amplifier, PC: polarization controller. \textbf{c} and \textbf{g},  Displacement and electrical field distributions at the waveguide cross-section for the quasi-Love phononic mode and the fundamental transverse electric photonic mode, respectively. \textbf{d} and \textbf{e}, Scanning electron microscope pictures of the IDT and phononic-photonic mode multiplexing device. \textbf{f}, S11 parameter of IDT which exhibits a resonance frequency of 8.5 GHz. \textbf{h}, Measured reflected optical signal against the input RF frequency, with the input optical wavelength fixed at $1576\,\mathrm{nm}$.} 
\label{Fig2}
\end{figure*}

\begin{figure*}[t]
\begin{centering}
\textcolor{red}{\includegraphics[width=1\textwidth]{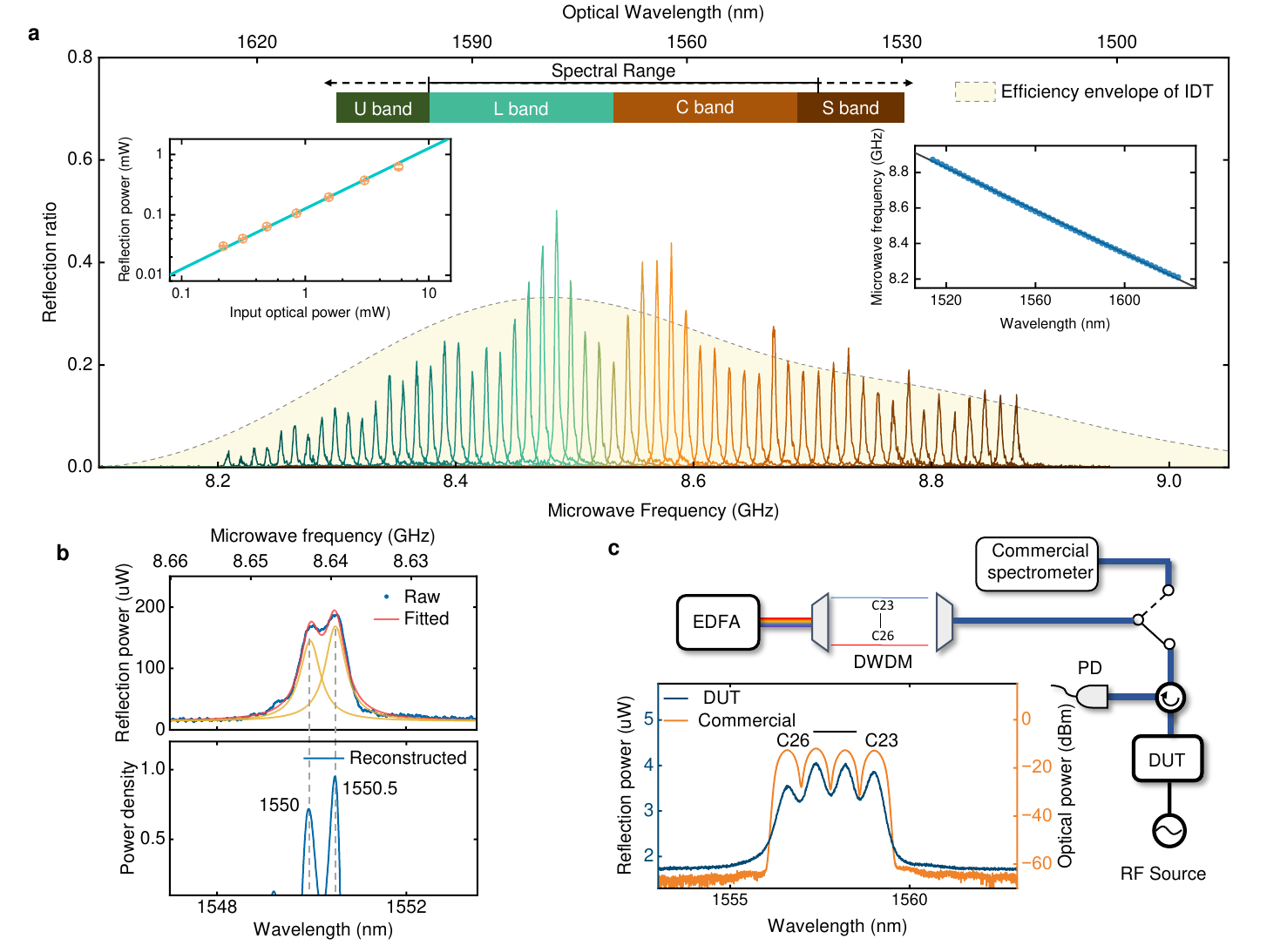}}
\par\end{centering}
\caption{\textbf{Broadband spectrum characterization.} \textbf{a}, Separate spectral line response with different pump wavelengths, with an RF drive power is $20\,\mathrm{mW}$. Inset:  the relationship between the reflected optical power and the input power; Dispersion relationship between the beat frequency of the output light and the input optical wavelength. The slope is $6.03\,\mathrm{MHz/nm}$. \textbf{b}, Top: RF response spectrum with two pump lasers at $1550.5\,\mathrm{nm}$ and $1550\,\mathrm{nm}$. The red line represents a double-Lorentz fitting for the measured data, which is the sum of two Lorentz distributions displayed in yellow. Bottom: Reconstructed optical spectrum. \textbf{c}, We use a standard commercial dense wavelength division multiplexing (DWDM) as a filter to generate the spectrum by tuning the spontaneous radiation of our EDFA without any laser input. The blue line represents the output signal from our device detected by a PD directly and the orange line comes from a commercial spectrometer.}
\label{Fig3}
\end{figure*}

\smallskip{}

\noindent \textbf{\large{}Results}{\large\par}

\noindent Figure~\ref{Fig1} illustrates the conceptual schematics of conventional integrated spectrometers and the principle of the proposed approach based on Brillouin scattering~\cite{eggleton2019brillouin,liu2019electromechanical}. Traditional on-chip spectrometers typically rely on linear optical processes, such as dispersive elements in Fig.~\ref{Fig1}a or Fourier-transform techniques in Fig.~\ref{Fig1}b. In contrast, our spectrometer exploits the nonlinear Brillouin interaction between guided photonic and phononic modes in a compact waveguide platform (Fig.~\ref{Fig1}c). At the heart of our device is a straight waveguide that supports both optical and acoustic modes, which are coupled primarily through the photoelastic effect. The waveguide is integrated with a compact interdigital transducer (IDT) for RF-to-acoustic wave conversion and with grating couplers that enable efficient excitation and detection of the optical response. When operating the spectrometer, an input optical signal ($\omega_1$) is coupled into the waveguide and interacts with a counter-propagating acoustic wave ($\Omega$) generated by the IDT. 

The Brillouin scattering process manifests as a coherent three-wave mixing (TWM) interaction, leading to a wavelength-dependent coupling between forward and backward optical signals stimulated by acoustic wave fields. This coherent TWM results in a frequency-shifted output signal ($\omega_2$) that satisfies both the energy and momentum conservation conditions: $\omega_2=\omega_1+\Omega$ and ${k_2}={k_1}+{\beta}$, where ${k_{1(2)}}$ and ${\beta}$ are the wavevectors of the input (output) optical signal and acoustic wave, respectively. In a simpler picture, the acoustic wave modulates the refractive index along the waveguide through the photoelastic effect, and thus the waveguide can be understood as a dynamic optical grating. Therefore, our spectrometer resembles a linear grating spectrometer (Fig.~\ref{Fig1}a) with the corresponding input and output angles $\theta_1=\theta_2=\pi/2$, allowing us to achieve a spectral resolution approaching the theoretical upper bound of ${\lambda^2}/{2n_{\mathrm{eff}}L}$ for a given interaction length $L$.  Moreover, our single-channel device can resolve a wide spectral range through dynamically reconfigurable ${\beta}$ by simply sweeping the external RF drive frequency.  

To experimentally validate our ultra-compact spectrometer, we fabricate the hybrid phononic and photonic circuits using thin-film LN on a sapphire substrate. The combination single-crystal LN and sapphire allows for low-loss photonic and phononic waveguides, while LN's strong piezoelectric and photoelastic properties facilitate efficient interfacing among the RF, acoustic, and optical signals, making it an ideal material platform for our spectrometer. The device, shown in Fig.~\ref{Fig2}b, features a straight waveguide with a width of $1.2\,\mathrm{\mu m}$, a thickness of $400\,\mathrm{nm}$, and a wedge step of $180\,\mathrm{nm}$. This carefully designed geometry supports both optical modes at telecom wavelengths and acoustic modes around $8.5\,\mathrm{GHz}$, with a phase-matched phonon wavelength of $400\,\mathrm{nm}$. The corresponding mode profiles of the photonic and phononic modes are shown in Fig.~\ref{Fig2}c and Fig.~\ref{Fig2}g. To excite and focus the acoustic wave into the waveguide, we employ a fan-shaped IDT (Fig.~\ref{Fig2}d) driven by a RF signal from a vector network analyzer (VNA).  The RF S11 parameter of the transducer, as shown in Fig.~\ref{Fig2}f, exhibits a resonance frequency of $8.5\,\mathrm{GHz}$, indicating the efficient generation of high-frequency guided phonons into the waveguide. 

To efficiently couple phonons and photons into the same waveguide for Brillouin interaction, while protecting the electrodes from potential damage caused by direct interaction between the input laser and the IDT, we have designed a hybrid phononic-photonic multiplexing device, as shown in Fig.~\ref{Fig2}e. This device exploits the tighter confinement of the phononic mode, allowing it to propagate through the waveguide with minimal coupling to adjacent waveguide due to the weak evanescent acoustic field. In contrast, the photonic mode can be efficiently transferred between the two waveguides when passing through the coupling region. Optical input and output couplings are achieved using grating couplers for high signal-to-noise ratio or using side couplings to avoid the sensitivity of coupling efficiency to wavelength and achieve a large wavelength bandwidth couple (A more detailed discussion provided in the Supplementary Information $\text{II}$.) 

The experimental setup for characterizing the Brillouin scattering is depicted in Fig.~\ref{Fig2}a. A tunable laser, amplified by an erbium-doped fiber amplifier (EDFA) and with its polarization controlled, is coupled into the waveguide. The reflected optical signal at the selected wavelength, which satisfies the phase-matching condition of Brillouin scattering in the spectrometer, is collected and directed to a photodetector. To investigate the spectrometer's performance, we fixed the input optical wavelength at $1576\,\mathrm{nm}$ and scanned the input RF signal supplied by the VNA, measuring the reflected light intensity with an interaction length $L=1\,\mathrm{mm}$. The result, shown in Fig.~\ref{Fig2}h, demonstrates the remarkably reflection of the acoustically-induced dynamic grating. The spectrum exhibits a narrow linewidth, with a full width at half maximum (FWHM) of just 3.4\,MHz, validating the highly selective spectral response and also sub-$\mu$s switching time for high-resolution on-chip spectral analysis. Moreover,  
with an RF drive power of only $20\,\mathrm{mW}$, as shown in Fig.~\ref{Fig3}a , the grating achieves a peak reflectivity of 50\%, which highlights the efficient Brillouin interaction enabled by the carefully engineered hybrid photonic-phononic waveguide.

Figure~\ref{Fig3} demonstrates the broadband performance of our spectrometer. In Fig.~\ref{Fig3}a, the responses for monochromatic optical input signals are measured for different wavelengths ranging from 1514 to 1624\,nm, with an experimental wavelength of 110\,nm limited by the spectral range of our testing optical source. The linear dependence between the input and reflected optical powers, as shown in the left inset of Fig.~\ref{Fig3}a, verifies the coherent TWM process of acoustically-stimulated Brillouin scattering. The right inset of Fig.~\ref{Fig3}a depicts the linear dispersion relationship between the frequency shift of the output light and the input optical wavelength, with a slope of 6.03\,MHz/nm. Combined with the narrow bandwidth of the RF response [Fig.~\ref{Fig2}h], we derive an optical intrinsic spectral resolution of 0.56\,nm. The envelope of the varying peak values represents the frequency-dependent RF-phonon conversion efficiency, which is determined by the bandwidth of the IDT. Although the experimental data is acquired over a 110\,nm range, our analysis (Supplementary Information) indicates that  the spectrometer can operate over a significantly broader wavelength range. Based on the characteristics of the current IDT design, we estimate that the dynamic range of the spectrometer exceeds 150\,nm. Furthermore, by employing advanced broadband IDT designs or cascaded IDTs on a single waveguide, the operating range can potentially be extended to 300\,nm or more.

To directly demonstrate the sub-nanometer resolution of our spectrometer, we perform a spectral measurement using a source consisting of two laser tones as input, with a wavelength interval of only 0.5\,nm. By scanning the RF drive frequency, we obtain the spectral information shown in Fig.~\ref{Fig3}b. The raw data clearly shows two distinctly resolved peaks. We fit the data using a double-Lorentzian function, which is the sum of two Lorentzian distributions displayed in red. Furthermore, by using the spectral response for monochromatic light in Fig.~\ref{Fig3}a, we reconstruct the input spectrum with respect to the optical wavelength. The retrieved wavelength data are consistent with our input laser wavelengths, proving that the projected resolution of our spectrometer can be much better than 0.5\,nm. 

To further demonstrate the fluorescence measurement capability of our spectrometer and its potential in telecom applications, we use an EDFA to generate a broadband spectrum and utilize a dense wavelength division multiplexing (DWDM) filter to synthesize a broadband light source. The DWDM filter allows us to selectively switch on and off specific bandpass channels (e.g., C23-C26) to create controlled spectral features. As shown in Fig.~\ref{Fig3}c, the spectrum is first measured using a commercial spectrometer as a reference and delivered to our device. In contrast to the heterodyne detection scheme mentioned in Fig.~\ref{Fig2}a, the output signal from our device is directly detected by a PD.  The response spectrum in Fig.~\ref{Fig3}c (blue line) reveals a frequency-independent background noise floor, which we attribute to reflections of the input signal in our system, primarily due to the backward scattering from the coupler. 
By mapping the RF response to the optical domain, we derive the optical spectrum, which exhibits excellent agreement with the reference spectrum measured by a commercial spectrometer (orange line). This measurement result demonstrates that our spectrometer possesses the capability to resolve fluorescence spectra and is well-suited for DWDM, which is widely used in the telecom technology.

\begin{figure}[t]
\begin{centering}
\textcolor{red}{\includegraphics[width=1\linewidth]{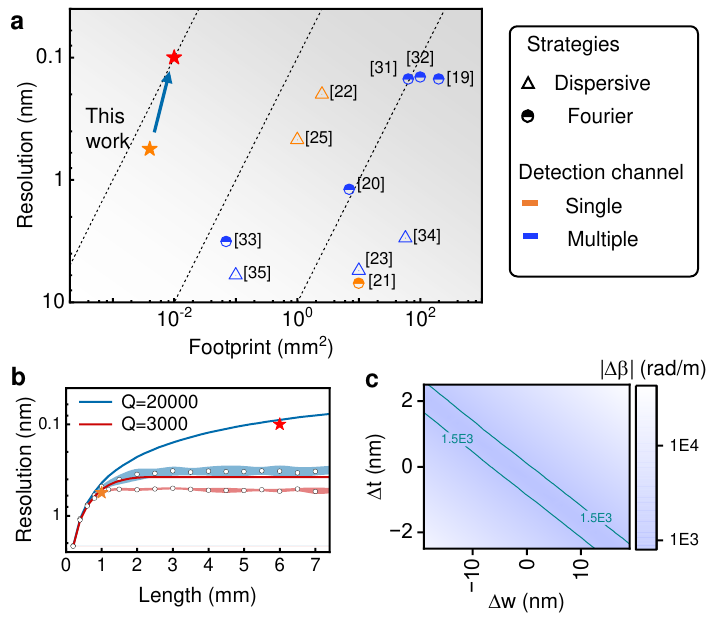}}
\par\end{centering}
\caption{\textbf{Spectrometer intrinsic resolution.} \textbf{a}, A plot comparing the resolution, detection channel, and footprint scale for selected device demonstrations in the literature. The footprint scale includes the elements in the device that are active in resolving and detecting light, excluding the accessory components. Reference numbers are indicated within square brackets. \textbf{b}, Simulation results of the device resolution performance improved by the interaction length ($L$). The red line represents the result based on the current material and fabrication level with a phonon Q $\sim 3000$. The blue line shows the resolution with an improvement of phonon Q $\sim 20000$ in future experiments. The dots represent the simulation results with the impact from the phase-matching diffusion and the shadow part represents the standard deviation of the resolution. The standard deviation of the random diffusion value added in simulation is about $1500\,\mathrm{rad/m}$, which represents the standard derivation of waveguide width  ($\Delta w$) around 4\,nm and film thickness ($\Delta t$) around 0.5\,nm. \textbf{c}, The heatmap represents the relationship between mismatching value $\Delta\beta$ and variation error on $\Delta w$ and $\Delta t$. The cyan line corresponds to the diffusion value in \textbf{b}.}
\label{Fig4}
\end{figure}

In Fig.~\ref{Fig4}a, we compare the performance of our device with state-of-the-art demonstrations reported in the literature~\cite{cheben2007high,koshelev2014combination,zou2016ultracompact,faraji2018compact,calafiore2014holographic,cheng2019broadband,nedeljkovic2015mid,pohl2020integrated,nie2017cmos,kita2018high,zheng2019microring}, focusing on the intrinsic spectral resolution of the devices. The intrinsic spectral resolution is governed by the dispersive path of the device, which is determined by its inherent optical response bandwidth, regardless of the input signal strength, detection noises, and the reconstruction algorithms~\cite{wang2019single,redding2013compact,yang2019single,he2024microsized,li2021chip}. Our device exhibits outstanding performance in terms of resolution and footprint. According to the coherent mode conversion in an ideal uniform waveguide, the intrinsic resolution is determined by the phase-matching condition in the coupled mode equations~\cite{sipe2016hamiltonian} (Supplementary Information). For a fixed RF frequency, the reflected light has a bandwidth corresponding to the FWHM of the wavevector $\Delta_\beta\approx{5.56}/{L}$, which implies a fundamental trade-off relation between resolution and footprint. Considering the dispersion relation of the photonic mode as ${\partial\beta}/{\partial\lambda}=1.2\times10^{4}\,\mathrm{rad}/(\mathrm{m}\cdot\mathrm{nm})$, the fundamental limitation of resolution is given by
\begin{equation}
\Delta_\lambda=\Delta_\beta/\frac{\partial\beta}{\partial\lambda}=\frac{460\,\mathrm{\mu m}}{L}\,\mathrm{nm}.    
\end{equation}
For an interaction length $L=1\,\mathrm{mm}$ in our device, the resolution limit for an ideal lossless waveguide is $0.46\,\mathrm{nm}$, indicating that our experimental result of $0.56\,\mathrm{nm}$ approaches this fundamental limit with a small discrepancy attributed to the losses. Also, we find that the practical resolution of our spectrometer does not improve with further increases in $L$, revealing the constraints imposed by the propagation loss of the phononic mode in the waveguide. As shown by the red lines in Fig.~\ref{Fig4}b, the resolution considering the phonon attenuation is calculated for given equivalent phononic quality factors ($Q$), and a $Q=3000$ explains our observations of the saturated resolution for $L>1\,\mathrm{mm}$. Furthermore, the potential imperfections of non-uniform waveguide cross-section can introduce variations in the mode wavevectors, which can be treated as a phase-matching diffusion process in the mode conversion and reduce the resolution. The dependence of the phase mismatching $\Delta\beta$ on the variation of the waveguide width ($\Delta w$) and film thickness ($\Delta t$) is shown in Fig.~\ref{Fig4}c, demonstrating that the two geometry parameters independently contribute to the mismatching. By assuming a random walk of $\Delta\beta$ and the accumulated mismatch in the interaction region following a normal distribution, we numerically simulate the influence of the phase-matching diffusion with a standard deviation $\sigma=1500\,\mathrm{rad/m}$ for different phonon $Q$ (Supplementary Information). The results are plotted in Fig.~\ref{Fig4}b as dots, with the shaded area representing the standard deviation. These results show that the impact of phase-matching diffusion becomes more pronounced for higher $Q$. Therefore, a resolution of $0.1\,\mathrm{nm}$ with an interaction length under $6\,\mathrm{mm}$ is feasible by improving phonon $Q$ by one order of magnitude~\cite{1981attenuation} and suppressing the variation of $\Delta\beta$ through compensating for film thickness variations via specially designed modulations in the waveguide width~\cite{linran2024adapted}.

\smallskip{}

\noindent \textbf{\large{}Discussion}{\large\par}

\noindent Our demonstration of the acoustically-stimulated Brillouin scattering introduces a new paradigm for on-chip spectroscopy by leveraging the unique properties of hybrid photonic-phononic circuits. The spectrometer presented in this work holds great promise for its simplicity, robustness, and ease of fabrication, as well as its potential for large-scale integration with other photonic and electronic components~\cite{elshaari2020hybrid,xu2022hybrid}. The absence of moving parts or suspended structures makes our device inherently stable and reliable, offering a compact, stable, and cost-effective solution for spectral analysis without the need for any free-space optical components. The performance of our spectrometer, particularly in terms of resolution and bandwidth, can be further enhanced by optimizing the device design and fabrication. 

Our work validates a hybrid photonic-phononic platform, which not only provides a novel approach for dispersion manipulation on a chip but also serves as the foundation for a new chip architecture. We term this platform as ``\textit{Zhengfu}" hybrid chip architecture, and it can be extended to other material platforms, such as scandium aluminum nitride (ScAlN) on sapphire~\cite{chen2022scandium} and lithium tantalate (LiTaO3) on sapphire~\cite{bartasyte2017toward}, offering flexibility in material choice  for specific applications. 
This architecture enables an efficient interface between RF and optical signals, opening up new possibilities for optoelectronic devices, and offering a complementary paradigm to electro-optic modulation for optical information processing. For instance, dynamically reconfigurable lasing on a chip can be realized by simply controlling the RF drive amplitude and frequency. This can enable fast frequency sweeping or multiple laser emission, paving the way for advanced on-chip laser sources with enhanced functionality and versatility. Moreover, the \textit{Zhengfu} architecture is compatible with superconducting quantum devices, such as SNSPD, superconducting resonators, and qubits, thereby presenting exciting opportunities for future quantum information processing. 

\smallskip{}

\noindent \textbf{\large{}Online content}{\large\par}

\noindent Any methods, additional references, Nature Research reporting summaries, source data, extended data, supplementary information, acknowledgements, peer review information; details of author contributions and competing interests; and statements of data and code availability are available online.

%


\noindent \textbf{\large{}Data availability}{\large\par}

\noindent All data generated or analysed during this study are available within the paper and its Supplementary Information. Further source data will be made available on reasonable request.

\smallskip{}

\noindent \textbf{\large{}Code availability}{\large\par}

\noindent The code used to solve the equations presented in the Supplementary Information will be made available on reasonable request.

\smallskip{}

\noindent \textbf{\large{}Acknowledgment}{\large\par}

\noindent We would like to express our gratitude to Prof. Yeteng Zhong for discussions and to Wen Liu for the assistance in fabrication. This work was funded by the National Key Research and Development Program (2017YFA0304303) and the National Natural Science Foundation of China (Grant Nos.~92265210, 123B2068, 12104441, 92165209, 12293053, 11925404, 12374361, and 92365301). We also acknowledge the support from the Fundamental Research Funds for the Central Universities and USTC Research Funds of the Double First-Class Initiative. The numerical calculations in this paper were performed on the supercomputing system in the Supercomputing Center of University of Science and Technology of China. This work was partially carried out at the USTC Center for Micro and Nanoscale Research and Fabrication.

\smallskip{}

\noindent \textbf{\large{}Author contributions}{\large\par}

\noindent C.-L.Z. conceived the experiments. J.-Q.W., Y.-H.Y. and Z.-X.Z. designed and fabricated the devices, built the experimental setup, carried out the measurements, and analyzed the data with the assistance from X.-B.X., L.S., Z.-X.Z., J.-J.L., and C.-H.D. M.L. provided theoretical supports. J.-Q.W. and C.-L.Z. wrote the manuscript with input from all other authors. X.-B.X., L.S. and C.-L.Z. supervised the project.

\smallskip{}

\noindent \textbf{\large{}Competing interests}{\large\par}

\noindent The authors declare no competing interests.

\smallskip{}

\noindent \textbf{\large{}Additional information}{\large\par}

\noindent \textbf{Supplementary information} The online version contains
supplementary material.

\noindent \textbf{Correspondence and requests for materials} should
be addressed to X.-B.X., L.S., and C.-L.Z.


\begin{thebibliography}{42}%
\makeatletter
\providecommand \@ifxundefined [1]{%
 \@ifx{#1\undefined}
}%
\providecommand \@ifnum [1]{%
 \ifnum #1\expandafter \@firstoftwo
 \else \expandafter \@secondoftwo
 \fi
}%
\providecommand \@ifx [1]{%
 \ifx #1\expandafter \@firstoftwo
 \else \expandafter \@secondoftwo
 \fi
}%
\providecommand \natexlab [1]{#1}%
\providecommand \enquote  [1]{``#1''}%
\providecommand \bibnamefont  [1]{#1}%
\providecommand \bibfnamefont [1]{#1}%
\providecommand \citenamefont [1]{#1}%
\providecommand \href@noop [0]{\@secondoftwo}%
\providecommand \href [0]{\begingroup \@sanitize@url \@href}%
\providecommand \@href[1]{\@@startlink{#1}\@@href}%
\providecommand \@@href[1]{\endgroup#1\@@endlink}%
\providecommand \@sanitize@url [0]{\catcode `\\12\catcode `\$12\catcode
  `\&12\catcode `\#12\catcode `\^12\catcode `\_12\catcode `\%12\relax}%
\providecommand \@@startlink[1]{}%
\providecommand \@@endlink[0]{}%
\providecommand \url  [0]{\begingroup\@sanitize@url \@url }%
\providecommand \@url [1]{\endgroup\@href {#1}{\urlprefix }}%
\providecommand \urlprefix  [0]{URL }%
\providecommand \Eprint [0]{\href }%
\providecommand \doibase [0]{http://dx.doi.org/}%
\providecommand \selectlanguage [0]{\@gobble}%
\providecommand \bibinfo  [0]{\@secondoftwo}%
\providecommand \bibfield  [0]{\@secondoftwo}%
\providecommand \translation [1]{[#1]}%
\providecommand \BibitemOpen [0]{}%
\providecommand \bibitemStop [0]{}%
\providecommand \bibitemNoStop [0]{.\EOS\space}%
\providecommand \EOS [0]{\spacefactor3000\relax}%
\providecommand \BibitemShut  [1]{\csname bibitem#1\endcsname}%
\let\auto@bib@innerbib\@empty
\bibitem [{\citenamefont {Elshaari}\ \emph {et~al.}(2020)\citenamefont
  {Elshaari}, \citenamefont {Pernice}, \citenamefont {Srinivasan},
  \citenamefont {Benson},\ and\ \citenamefont {Zwiller}}]{elshaari2020hybrid}%
  \BibitemOpen
  \bibfield  {author} {\bibinfo {author} {\bibfnamefont {A.~W.}\ \bibnamefont
  {Elshaari}}, \bibinfo {author} {\bibfnamefont {W.}~\bibnamefont {Pernice}},
  \bibinfo {author} {\bibfnamefont {K.}~\bibnamefont {Srinivasan}}, \bibinfo
  {author} {\bibfnamefont {O.}~\bibnamefont {Benson}}, \ and\ \bibinfo {author}
  {\bibfnamefont {V.}~\bibnamefont {Zwiller}},\ }\bibfield  {title} {\enquote
  {\bibinfo {title} {Hybrid integrated quantum photonic circuits},}\ }\href
  {\doibase 10.1038/s41566-020-0609-x} {\bibfield  {journal} {\bibinfo
  {journal} {Nat. Photon.}\ }\textbf {\bibinfo {volume} {14}},\ \bibinfo
  {pages} {285} (\bibinfo {year} {2020})}\BibitemShut {NoStop}%
\bibitem [{\citenamefont {Wang}\ \emph {et~al.}(2020)\citenamefont {Wang},
  \citenamefont {Sciarrino}, \citenamefont {Laing},\ and\ \citenamefont
  {Thompson}}]{wang2020integrated}%
  \BibitemOpen
  \bibfield  {author} {\bibinfo {author} {\bibfnamefont {J.}~\bibnamefont
  {Wang}}, \bibinfo {author} {\bibfnamefont {F.}~\bibnamefont {Sciarrino}},
  \bibinfo {author} {\bibfnamefont {A.}~\bibnamefont {Laing}}, \ and\ \bibinfo
  {author} {\bibfnamefont {M.~G.}\ \bibnamefont {Thompson}},\ }\bibfield
  {title} {\enquote {\bibinfo {title} {Integrated photonic quantum
  technologies},}\ }\href {\doibase 10.1038/s41566-019-0532-1} {\bibfield
  {journal} {\bibinfo  {journal} {Nat. Photon.}\ }\textbf {\bibinfo {volume}
  {14}},\ \bibinfo {pages} {273} (\bibinfo {year} {2020})}\BibitemShut
  {NoStop}%
\bibitem [{\citenamefont {Shekhar}\ \emph {et~al.}(2024)\citenamefont
  {Shekhar}, \citenamefont {Bogaerts}, \citenamefont {Chrostowski},
  \citenamefont {Bowers}, \citenamefont {Hochberg}, \citenamefont {Soref},\
  and\ \citenamefont {Shastri}}]{Shekhar2024}%
  \BibitemOpen
  \bibfield  {author} {\bibinfo {author} {\bibfnamefont {S.}~\bibnamefont
  {Shekhar}}, \bibinfo {author} {\bibfnamefont {W.}~\bibnamefont {Bogaerts}},
  \bibinfo {author} {\bibfnamefont {L.}~\bibnamefont {Chrostowski}}, \bibinfo
  {author} {\bibfnamefont {J.~E.}\ \bibnamefont {Bowers}}, \bibinfo {author}
  {\bibfnamefont {M.}~\bibnamefont {Hochberg}}, \bibinfo {author}
  {\bibfnamefont {R.}~\bibnamefont {Soref}}, \ and\ \bibinfo {author}
  {\bibfnamefont {B.~J.}\ \bibnamefont {Shastri}},\ }\bibfield  {title}
  {\enquote {\bibinfo {title} {{Roadmapping the next generation of silicon
  photonics}},}\ }\href {\doibase 10.1038/s41467-024-44750-0} {\bibfield
  {journal} {\bibinfo  {journal} {Nat. Commun.}\ }\textbf {\bibinfo {volume}
  {15}},\ \bibinfo {pages} {751} (\bibinfo {year} {2024})}\BibitemShut
  {NoStop}%
\bibitem [{\citenamefont {Pelucchi}\ \emph {et~al.}(2022)\citenamefont
  {Pelucchi}, \citenamefont {Fagas}, \citenamefont {Aharonovich}, \citenamefont
  {Englund}, \citenamefont {Figueroa}, \citenamefont {Gong}, \citenamefont
  {Hannes}, \citenamefont {Liu}, \citenamefont {Lu}, \citenamefont {Matsuda}
  \emph {et~al.}}]{pelucchi2022potential}%
  \BibitemOpen
  \bibfield  {author} {\bibinfo {author} {\bibfnamefont {E.}~\bibnamefont
  {Pelucchi}}, \bibinfo {author} {\bibfnamefont {G.}~\bibnamefont {Fagas}},
  \bibinfo {author} {\bibfnamefont {I.}~\bibnamefont {Aharonovich}}, \bibinfo
  {author} {\bibfnamefont {D.}~\bibnamefont {Englund}}, \bibinfo {author}
  {\bibfnamefont {E.}~\bibnamefont {Figueroa}}, \bibinfo {author}
  {\bibfnamefont {Q.}~\bibnamefont {Gong}}, \bibinfo {author} {\bibfnamefont
  {H.}~\bibnamefont {Hannes}}, \bibinfo {author} {\bibfnamefont
  {J.}~\bibnamefont {Liu}}, \bibinfo {author} {\bibfnamefont {C.-Y.}\
  \bibnamefont {Lu}}, \bibinfo {author} {\bibfnamefont {N.}~\bibnamefont
  {Matsuda}},  \emph {et~al.},\ }\bibfield  {title} {\enquote {\bibinfo {title}
  {The potential and global outlook of integrated photonics for quantum
  technologies},}\ }\href {\doibase https://doi.org/10.1038/s42254-021-00398-z}
  {\bibfield  {journal} {\bibinfo  {journal} {Nat. Rev. Phys.}\ }\textbf
  {\bibinfo {volume} {4}},\ \bibinfo {pages} {194} (\bibinfo {year}
  {2022})}\BibitemShut {NoStop}%
\bibitem [{\citenamefont {Kippenberg}\ and\ \citenamefont
  {Vahala}(2008)}]{kippenberg2008cavity}%
  \BibitemOpen
  \bibfield  {author} {\bibinfo {author} {\bibfnamefont {T.~J.}\ \bibnamefont
  {Kippenberg}}\ and\ \bibinfo {author} {\bibfnamefont {K.~J.}\ \bibnamefont
  {Vahala}},\ }\bibfield  {title} {\enquote {\bibinfo {title} {Cavity
  optomechanics: back-action at the mesoscale},}\ }\href {\doibase
  10.1126/science.1156032} {\bibfield  {journal} {\bibinfo  {journal}
  {Science}\ }\textbf {\bibinfo {volume} {321}},\ \bibinfo {pages} {1172}
  (\bibinfo {year} {2008})}\BibitemShut {NoStop}%
\bibitem [{\citenamefont {Wang}\ \emph {et~al.}(2018)\citenamefont {Wang},
  \citenamefont {Zhang}, \citenamefont {Chen}, \citenamefont {Bertrand},
  \citenamefont {Shams-Ansari}, \citenamefont {Chandrasekhar}, \citenamefont
  {Winzer},\ and\ \citenamefont {Lon{\v{c}}ar}}]{wang2018integrated}%
  \BibitemOpen
  \bibfield  {author} {\bibinfo {author} {\bibfnamefont {C.}~\bibnamefont
  {Wang}}, \bibinfo {author} {\bibfnamefont {M.}~\bibnamefont {Zhang}},
  \bibinfo {author} {\bibfnamefont {X.}~\bibnamefont {Chen}}, \bibinfo {author}
  {\bibfnamefont {M.}~\bibnamefont {Bertrand}}, \bibinfo {author}
  {\bibfnamefont {A.}~\bibnamefont {Shams-Ansari}}, \bibinfo {author}
  {\bibfnamefont {S.}~\bibnamefont {Chandrasekhar}}, \bibinfo {author}
  {\bibfnamefont {P.}~\bibnamefont {Winzer}}, \ and\ \bibinfo {author}
  {\bibfnamefont {M.}~\bibnamefont {Lon{\v{c}}ar}},\ }\bibfield  {title}
  {\enquote {\bibinfo {title} {Integrated lithium niobate electro-optic
  modulators operating at cmos-compatible voltages},}\ }\href {\doibase
  https://doi.org/10.1038/s41586-018-0551-y} {\bibfield  {journal} {\bibinfo
  {journal} {Nature}\ }\textbf {\bibinfo {volume} {562}},\ \bibinfo {pages}
  {101} (\bibinfo {year} {2018})}\BibitemShut {NoStop}%
\bibitem [{\citenamefont {Corcoran}\ \emph {et~al.}(2020)\citenamefont
  {Corcoran}, \citenamefont {Tan}, \citenamefont {Xu}, \citenamefont {Boes},
  \citenamefont {Wu}, \citenamefont {Nguyen}, \citenamefont {Chu},
  \citenamefont {Little}, \citenamefont {Morandotti}, \citenamefont {Mitchell}
  \emph {et~al.}}]{corcoran2020ultra}%
  \BibitemOpen
  \bibfield  {author} {\bibinfo {author} {\bibfnamefont {B.}~\bibnamefont
  {Corcoran}}, \bibinfo {author} {\bibfnamefont {M.}~\bibnamefont {Tan}},
  \bibinfo {author} {\bibfnamefont {X.}~\bibnamefont {Xu}}, \bibinfo {author}
  {\bibfnamefont {A.}~\bibnamefont {Boes}}, \bibinfo {author} {\bibfnamefont
  {J.}~\bibnamefont {Wu}}, \bibinfo {author} {\bibfnamefont {T.~G.}\
  \bibnamefont {Nguyen}}, \bibinfo {author} {\bibfnamefont {S.~T.}\
  \bibnamefont {Chu}}, \bibinfo {author} {\bibfnamefont {B.~E.}\ \bibnamefont
  {Little}}, \bibinfo {author} {\bibfnamefont {R.}~\bibnamefont {Morandotti}},
  \bibinfo {author} {\bibfnamefont {A.}~\bibnamefont {Mitchell}},  \emph
  {et~al.},\ }\bibfield  {title} {\enquote {\bibinfo {title} {Ultra-dense
  optical data transmission over standard fibre with a single chip source},}\
  }\href {\doibase 10.1038/s41467-020-16265-x} {\bibfield  {journal} {\bibinfo
  {journal} {Nat. Commun.}\ }\textbf {\bibinfo {volume} {11}},\ \bibinfo
  {pages} {2568} (\bibinfo {year} {2020})}\BibitemShut {NoStop}%
\bibitem [{\citenamefont {Shen}\ \emph {et~al.}(2014)\citenamefont {Shen},
  \citenamefont {Li},\ and\ \citenamefont {Li}}]{shen2014silicon}%
  \BibitemOpen
  \bibfield  {author} {\bibinfo {author} {\bibfnamefont {M.-Y.}\ \bibnamefont
  {Shen}}, \bibinfo {author} {\bibfnamefont {B.-R.}\ \bibnamefont {Li}}, \ and\
  \bibinfo {author} {\bibfnamefont {Y.-K.}\ \bibnamefont {Li}},\ }\bibfield
  {title} {\enquote {\bibinfo {title} {Silicon nanowire field-effect-transistor
  based biosensors: From sensitive to ultra-sensitive},}\ }\href {\doibase
  https://doi.org/10.1016/j.bios.2014.03.057} {\bibfield  {journal} {\bibinfo
  {journal} {Biosens.Bioelectron.}\ }\textbf {\bibinfo {volume} {60}},\
  \bibinfo {pages} {101} (\bibinfo {year} {2014})}\BibitemShut {NoStop}%
\bibitem [{\citenamefont {Eggleton}\ \emph {et~al.}(2019)\citenamefont
  {Eggleton}, \citenamefont {Poulton}, \citenamefont {Rakich}, \citenamefont
  {Steel},\ and\ \citenamefont {Bahl}}]{eggleton2019brillouin}%
  \BibitemOpen
  \bibfield  {author} {\bibinfo {author} {\bibfnamefont {B.~J.}\ \bibnamefont
  {Eggleton}}, \bibinfo {author} {\bibfnamefont {C.~G.}\ \bibnamefont
  {Poulton}}, \bibinfo {author} {\bibfnamefont {P.~T.}\ \bibnamefont {Rakich}},
  \bibinfo {author} {\bibfnamefont {M.~J.}\ \bibnamefont {Steel}}, \ and\
  \bibinfo {author} {\bibfnamefont {G.}~\bibnamefont {Bahl}},\ }\bibfield
  {title} {\enquote {\bibinfo {title} {Brillouin integrated photonics},}\
  }\href {\doibase 10.1038/s41566-019-0498-z} {\bibfield  {journal} {\bibinfo
  {journal} {Nat. Photon.}\ }\textbf {\bibinfo {volume} {13}},\ \bibinfo
  {pages} {664} (\bibinfo {year} {2019})}\BibitemShut {NoStop}%
\bibitem [{\citenamefont {Jovanovic}\ \emph {et~al.}(2023)\citenamefont
  {Jovanovic}, \citenamefont {Gatkine}, \citenamefont {Anugu}, \citenamefont
  {Amezcua-Correa}, \citenamefont {Thakur}, \citenamefont {Beichman},
  \citenamefont {Bender}, \citenamefont {Berger}, \citenamefont {Bigioli},
  \citenamefont {Bland-Hawthorn} \emph {et~al.}}]{jovanovic2023}%
  \BibitemOpen
  \bibfield  {author} {\bibinfo {author} {\bibfnamefont {N.}~\bibnamefont
  {Jovanovic}}, \bibinfo {author} {\bibfnamefont {P.}~\bibnamefont {Gatkine}},
  \bibinfo {author} {\bibfnamefont {N.}~\bibnamefont {Anugu}}, \bibinfo
  {author} {\bibfnamefont {R.}~\bibnamefont {Amezcua-Correa}}, \bibinfo
  {author} {\bibfnamefont {R.~B.}\ \bibnamefont {Thakur}}, \bibinfo {author}
  {\bibfnamefont {C.}~\bibnamefont {Beichman}}, \bibinfo {author}
  {\bibfnamefont {C.~F.}\ \bibnamefont {Bender}}, \bibinfo {author}
  {\bibfnamefont {J.-P.}\ \bibnamefont {Berger}}, \bibinfo {author}
  {\bibfnamefont {A.}~\bibnamefont {Bigioli}}, \bibinfo {author} {\bibfnamefont
  {J.}~\bibnamefont {Bland-Hawthorn}},  \emph {et~al.},\ }\bibfield  {title}
  {\enquote {\bibinfo {title} {2023 astrophotonics roadmap: pathways to
  realizing multi-functional integrated astrophotonic instruments},}\ }\href
  {\doibase https://doi.org/10.1088/2515-7647/ace869} {\bibfield  {journal}
  {\bibinfo  {journal} {New J. Phys.: Photon.}\ }\textbf {\bibinfo {volume}
  {5}},\ \bibinfo {pages} {042501} (\bibinfo {year} {2023})}\BibitemShut
  {NoStop}%
\bibitem [{\citenamefont {Yu}\ \emph {et~al.}(2018)\citenamefont {Yu},
  \citenamefont {Okawachi}, \citenamefont {Griffith}, \citenamefont
  {Picqu{\'e}}, \citenamefont {Lipson},\ and\ \citenamefont
  {Gaeta}}]{yu2018silicon}%
  \BibitemOpen
  \bibfield  {author} {\bibinfo {author} {\bibfnamefont {M.}~\bibnamefont
  {Yu}}, \bibinfo {author} {\bibfnamefont {Y.}~\bibnamefont {Okawachi}},
  \bibinfo {author} {\bibfnamefont {A.~G.}\ \bibnamefont {Griffith}}, \bibinfo
  {author} {\bibfnamefont {N.}~\bibnamefont {Picqu{\'e}}}, \bibinfo {author}
  {\bibfnamefont {M.}~\bibnamefont {Lipson}}, \ and\ \bibinfo {author}
  {\bibfnamefont {A.~L.}\ \bibnamefont {Gaeta}},\ }\bibfield  {title} {\enquote
  {\bibinfo {title} {Silicon-chip-based mid-infrared dual-comb spectroscopy},}\
  }\href {\doibase https://doi.org/10.1038/s41467-018-04350-1} {\bibfield
  {journal} {\bibinfo  {journal} {Nat. Commun.}\ }\textbf {\bibinfo {volume}
  {9}},\ \bibinfo {pages} {1869} (\bibinfo {year} {2018})}\BibitemShut
  {NoStop}%
\bibitem [{\citenamefont {Kong}\ \emph {et~al.}(2015)\citenamefont {Kong},
  \citenamefont {Kendall}, \citenamefont {Stone},\ and\ \citenamefont
  {Notingher}}]{kong2015raman}%
  \BibitemOpen
  \bibfield  {author} {\bibinfo {author} {\bibfnamefont {K.}~\bibnamefont
  {Kong}}, \bibinfo {author} {\bibfnamefont {C.}~\bibnamefont {Kendall}},
  \bibinfo {author} {\bibfnamefont {N.}~\bibnamefont {Stone}}, \ and\ \bibinfo
  {author} {\bibfnamefont {I.}~\bibnamefont {Notingher}},\ }\bibfield  {title}
  {\enquote {\bibinfo {title} {Raman spectroscopy for medical diagnostics-from
  in-vitro biofluid assays to in-vivo cancer detection},}\ }\href {\doibase
  10.1016/j.addr.2015.03.009} {\bibfield  {journal} {\bibinfo  {journal} {Adv.
  Drug Deliv. Rev.}\ }\textbf {\bibinfo {volume} {89}},\ \bibinfo {pages} {121}
  (\bibinfo {year} {2015})}\BibitemShut {NoStop}%
\bibitem [{\citenamefont {Ho}\ \emph {et~al.}(2005)\citenamefont {Ho},
  \citenamefont {Robinson}, \citenamefont {Miller},\ and\ \citenamefont
  {Davis}}]{ho2005overview}%
  \BibitemOpen
  \bibfield  {author} {\bibinfo {author} {\bibfnamefont {C.~K.}\ \bibnamefont
  {Ho}}, \bibinfo {author} {\bibfnamefont {A.}~\bibnamefont {Robinson}},
  \bibinfo {author} {\bibfnamefont {D.~R.}\ \bibnamefont {Miller}}, \ and\
  \bibinfo {author} {\bibfnamefont {M.~J.}\ \bibnamefont {Davis}},\ }\bibfield
  {title} {\enquote {\bibinfo {title} {Overview of sensors and needs for
  environmental monitoring},}\ }\href {\doibase
  https://doi.org/10.3390/s5010004} {\bibfield  {journal} {\bibinfo  {journal}
  {Sensors}\ }\textbf {\bibinfo {volume} {5}},\ \bibinfo {pages} {4} (\bibinfo
  {year} {2005})}\BibitemShut {NoStop}%
\bibitem [{\citenamefont {Kneipp}\ \emph {et~al.}(1999)\citenamefont {Kneipp},
  \citenamefont {Kneipp}, \citenamefont {Itzkan}, \citenamefont {Dasari},\ and\
  \citenamefont {Feld}}]{kneipp1999ultrasensitive}%
  \BibitemOpen
  \bibfield  {author} {\bibinfo {author} {\bibfnamefont {K.}~\bibnamefont
  {Kneipp}}, \bibinfo {author} {\bibfnamefont {H.}~\bibnamefont {Kneipp}},
  \bibinfo {author} {\bibfnamefont {I.}~\bibnamefont {Itzkan}}, \bibinfo
  {author} {\bibfnamefont {R.~R.}\ \bibnamefont {Dasari}}, \ and\ \bibinfo
  {author} {\bibfnamefont {M.~S.}\ \bibnamefont {Feld}},\ }\bibfield  {title}
  {\enquote {\bibinfo {title} {Ultrasensitive chemical analysis by raman
  spectroscopy},}\ }\href {\doibase https://doi.org/10.1021/cr980133r}
  {\bibfield  {journal} {\bibinfo  {journal} {Chem. Rev.}\ }\textbf {\bibinfo
  {volume} {99}},\ \bibinfo {pages} {2957} (\bibinfo {year}
  {1999})}\BibitemShut {NoStop}%
\bibitem [{\citenamefont {Harper}\ \emph {et~al.}(2024)\citenamefont {Harper},
  \citenamefont {Hwang}, \citenamefont {Kocheril}, \citenamefont {Lam},\ and\
  \citenamefont {Cushing}}]{Harper24}%
  \BibitemOpen
  \bibfield  {author} {\bibinfo {author} {\bibfnamefont {N.~A.}\ \bibnamefont
  {Harper}}, \bibinfo {author} {\bibfnamefont {E.~Y.}\ \bibnamefont {Hwang}},
  \bibinfo {author} {\bibfnamefont {P.~A.}\ \bibnamefont {Kocheril}}, \bibinfo
  {author} {\bibfnamefont {T.~K.}\ \bibnamefont {Lam}}, \ and\ \bibinfo
  {author} {\bibfnamefont {S.~K.}\ \bibnamefont {Cushing}},\ }\bibfield
  {title} {\enquote {\bibinfo {title} {Subtleties of nanophotonic lithium
  niobate waveguides for on-chip evanescent wave sensing},}\ }\href {\doibase
  10.1364/OE.529570} {\bibfield  {journal} {\bibinfo  {journal} {Opt. Express}\
  }\textbf {\bibinfo {volume} {32}},\ \bibinfo {pages} {27931} (\bibinfo {year}
  {2024})}\BibitemShut {NoStop}%
\bibitem [{\citenamefont {Yang}\ \emph {et~al.}(2021)\citenamefont {Yang},
  \citenamefont {Albrow-Owen}, \citenamefont {Cai},\ and\ \citenamefont
  {Hasan}}]{science2021miniaturization}%
  \BibitemOpen
  \bibfield  {author} {\bibinfo {author} {\bibfnamefont {Z.}~\bibnamefont
  {Yang}}, \bibinfo {author} {\bibfnamefont {T.}~\bibnamefont {Albrow-Owen}},
  \bibinfo {author} {\bibfnamefont {W.}~\bibnamefont {Cai}}, \ and\ \bibinfo
  {author} {\bibfnamefont {T.}~\bibnamefont {Hasan}},\ }\bibfield  {title}
  {\enquote {\bibinfo {title} {Miniaturization of optical spectrometers},}\
  }\href {\doibase 10.1126/science.abe0722} {\bibfield  {journal} {\bibinfo
  {journal} {Science}\ }\textbf {\bibinfo {volume} {371}},\ \bibinfo {pages}
  {eabe0722} (\bibinfo {year} {2021})}\BibitemShut {NoStop}%
\bibitem [{\citenamefont {Bacon}\ \emph {et~al.}(2004)\citenamefont {Bacon},
  \citenamefont {Mattley},\ and\ \citenamefont {DeFrece}}]{bacon2004miniature}%
  \BibitemOpen
  \bibfield  {author} {\bibinfo {author} {\bibfnamefont {C.~P.}\ \bibnamefont
  {Bacon}}, \bibinfo {author} {\bibfnamefont {Y.}~\bibnamefont {Mattley}}, \
  and\ \bibinfo {author} {\bibfnamefont {R.}~\bibnamefont {DeFrece}},\
  }\bibfield  {title} {\enquote {\bibinfo {title} {Miniature spectroscopic
  instrumentation: applications to biology and chemistry},}\ }\href {\doibase
  https://doi.org/10.1063/1.1633025} {\bibfield  {journal} {\bibinfo  {journal}
  {Rev. Sci. Instrum.}\ }\textbf {\bibinfo {volume} {75}},\ \bibinfo {pages}
  {1} (\bibinfo {year} {2004})}\BibitemShut {NoStop}%
\bibitem [{\citenamefont {Crocombe}(2018)}]{crocombe2018portable}%
  \BibitemOpen
  \bibfield  {author} {\bibinfo {author} {\bibfnamefont {R.~A.}\ \bibnamefont
  {Crocombe}},\ }\bibfield  {title} {\enquote {\bibinfo {title} {Portable
  spectroscopy},}\ }\href {\doibase https://doi.org/10.1177/0003702818809719}
  {\bibfield  {journal} {\bibinfo  {journal} {Appl. Spectrosc.}\ }\textbf
  {\bibinfo {volume} {72}},\ \bibinfo {pages} {1701} (\bibinfo {year}
  {2018})}\BibitemShut {NoStop}%
\bibitem [{\citenamefont {Calafiore}\ \emph {et~al.}(2014)\citenamefont
  {Calafiore}, \citenamefont {Koshelev}, \citenamefont {Dhuey}, \citenamefont
  {Goltsov}, \citenamefont {Sasorov}, \citenamefont {Babin}, \citenamefont
  {Yankov}, \citenamefont {Cabrini},\ and\ \citenamefont
  {Peroz}}]{calafiore2014holographic}%
  \BibitemOpen
  \bibfield  {author} {\bibinfo {author} {\bibfnamefont {G.}~\bibnamefont
  {Calafiore}}, \bibinfo {author} {\bibfnamefont {A.}~\bibnamefont {Koshelev}},
  \bibinfo {author} {\bibfnamefont {S.}~\bibnamefont {Dhuey}}, \bibinfo
  {author} {\bibfnamefont {A.}~\bibnamefont {Goltsov}}, \bibinfo {author}
  {\bibfnamefont {P.}~\bibnamefont {Sasorov}}, \bibinfo {author} {\bibfnamefont
  {S.}~\bibnamefont {Babin}}, \bibinfo {author} {\bibfnamefont
  {V.}~\bibnamefont {Yankov}}, \bibinfo {author} {\bibfnamefont
  {S.}~\bibnamefont {Cabrini}}, \ and\ \bibinfo {author} {\bibfnamefont
  {C.}~\bibnamefont {Peroz}},\ }\bibfield  {title} {\enquote {\bibinfo {title}
  {Holographic planar lightwave circuit for on-chip spectroscopy},}\ }\href
  {\doibase https://doi.org/10.1038/lsa.2014.84} {\bibfield  {journal}
  {\bibinfo  {journal} {Light Sci. Appl.}\ }\textbf {\bibinfo {volume} {3}},\
  \bibinfo {pages} {e203} (\bibinfo {year} {2014})}\BibitemShut {NoStop}%
\bibitem [{\citenamefont {Faraji-Dana}\ \emph {et~al.}(2018)\citenamefont
  {Faraji-Dana}, \citenamefont {Arbabi}, \citenamefont {Arbabi}, \citenamefont
  {Kamali}, \citenamefont {Kwon},\ and\ \citenamefont
  {Faraon}}]{faraji2018compact}%
  \BibitemOpen
  \bibfield  {author} {\bibinfo {author} {\bibfnamefont {M.}~\bibnamefont
  {Faraji-Dana}}, \bibinfo {author} {\bibfnamefont {E.}~\bibnamefont {Arbabi}},
  \bibinfo {author} {\bibfnamefont {A.}~\bibnamefont {Arbabi}}, \bibinfo
  {author} {\bibfnamefont {S.~M.}\ \bibnamefont {Kamali}}, \bibinfo {author}
  {\bibfnamefont {H.}~\bibnamefont {Kwon}}, \ and\ \bibinfo {author}
  {\bibfnamefont {A.}~\bibnamefont {Faraon}},\ }\bibfield  {title} {\enquote
  {\bibinfo {title} {Compact folded metasurface spectrometer},}\ }\href
  {\doibase https://doi.org/10.1038/s41467-018-06495-5} {\bibfield  {journal}
  {\bibinfo  {journal} {Nat. Commun.}\ }\textbf {\bibinfo {volume} {9}},\
  \bibinfo {pages} {4196} (\bibinfo {year} {2018})}\BibitemShut {NoStop}%
\bibitem [{\citenamefont {Cheng}\ \emph {et~al.}(2019)\citenamefont {Cheng},
  \citenamefont {Zou}, \citenamefont {Guo}, \citenamefont {Wang}, \citenamefont
  {Han},\ and\ \citenamefont {Tang}}]{cheng2019broadband}%
  \BibitemOpen
  \bibfield  {author} {\bibinfo {author} {\bibfnamefont {R.}~\bibnamefont
  {Cheng}}, \bibinfo {author} {\bibfnamefont {C.-L.}\ \bibnamefont {Zou}},
  \bibinfo {author} {\bibfnamefont {X.}~\bibnamefont {Guo}}, \bibinfo {author}
  {\bibfnamefont {S.}~\bibnamefont {Wang}}, \bibinfo {author} {\bibfnamefont
  {X.}~\bibnamefont {Han}}, \ and\ \bibinfo {author} {\bibfnamefont {H.~X.}\
  \bibnamefont {Tang}},\ }\bibfield  {title} {\enquote {\bibinfo {title}
  {Broadband on-chip single-photon spectrometer},}\ }\href {\doibase
  10.1038/s41467-019-12149-x} {\bibfield  {journal} {\bibinfo  {journal} {Nat.
  Commun.}\ }\textbf {\bibinfo {volume} {10}},\ \bibinfo {pages} {4104}
  (\bibinfo {year} {2019})}\BibitemShut {NoStop}%
\bibitem [{\citenamefont {Kita}\ \emph {et~al.}(2018)\citenamefont {Kita},
  \citenamefont {Miranda}, \citenamefont {Favela}, \citenamefont {Bono},
  \citenamefont {Michon}, \citenamefont {Lin}, \citenamefont {Gu},\ and\
  \citenamefont {Hu}}]{kita2018high}%
  \BibitemOpen
  \bibfield  {author} {\bibinfo {author} {\bibfnamefont {D.~M.}\ \bibnamefont
  {Kita}}, \bibinfo {author} {\bibfnamefont {B.}~\bibnamefont {Miranda}},
  \bibinfo {author} {\bibfnamefont {D.}~\bibnamefont {Favela}}, \bibinfo
  {author} {\bibfnamefont {D.}~\bibnamefont {Bono}}, \bibinfo {author}
  {\bibfnamefont {J.}~\bibnamefont {Michon}}, \bibinfo {author} {\bibfnamefont
  {H.}~\bibnamefont {Lin}}, \bibinfo {author} {\bibfnamefont {T.}~\bibnamefont
  {Gu}}, \ and\ \bibinfo {author} {\bibfnamefont {J.}~\bibnamefont {Hu}},\
  }\bibfield  {title} {\enquote {\bibinfo {title} {High-performance and
  scalable on-chip digital fourier transform spectroscopy},}\ }\href {\doibase
  https://doi.org/10.1038/s41467-018-06773-2} {\bibfield  {journal} {\bibinfo
  {journal} {Nat. Commun.}\ }\textbf {\bibinfo {volume} {9}},\ \bibinfo {pages}
  {4405} (\bibinfo {year} {2018})}\BibitemShut {NoStop}%
\bibitem [{\citenamefont {Pohl}\ \emph {et~al.}(2020)\citenamefont {Pohl},
  \citenamefont {Reig~Escal{\'e}}, \citenamefont {Madi}, \citenamefont
  {Kaufmann}, \citenamefont {Brotzer}, \citenamefont {Sergeyev}, \citenamefont
  {Guldimann}, \citenamefont {Giaccari}, \citenamefont {Alberti}, \citenamefont
  {Meier} \emph {et~al.}}]{pohl2020integrated}%
  \BibitemOpen
  \bibfield  {author} {\bibinfo {author} {\bibfnamefont {D.}~\bibnamefont
  {Pohl}}, \bibinfo {author} {\bibfnamefont {M.}~\bibnamefont
  {Reig~Escal{\'e}}}, \bibinfo {author} {\bibfnamefont {M.}~\bibnamefont
  {Madi}}, \bibinfo {author} {\bibfnamefont {F.}~\bibnamefont {Kaufmann}},
  \bibinfo {author} {\bibfnamefont {P.}~\bibnamefont {Brotzer}}, \bibinfo
  {author} {\bibfnamefont {A.}~\bibnamefont {Sergeyev}}, \bibinfo {author}
  {\bibfnamefont {B.}~\bibnamefont {Guldimann}}, \bibinfo {author}
  {\bibfnamefont {P.}~\bibnamefont {Giaccari}}, \bibinfo {author}
  {\bibfnamefont {E.}~\bibnamefont {Alberti}}, \bibinfo {author} {\bibfnamefont
  {U.}~\bibnamefont {Meier}},  \emph {et~al.},\ }\bibfield  {title} {\enquote
  {\bibinfo {title} {An integrated broadband spectrometer on thin-film lithium
  niobate},}\ }\href {\doibase https://doi.org/10.1038/s41566-019-0529-9}
  {\bibfield  {journal} {\bibinfo  {journal} {Nat Photon.}\ }\textbf {\bibinfo
  {volume} {14}},\ \bibinfo {pages} {24} (\bibinfo {year} {2020})}\BibitemShut
  {NoStop}%
\bibitem [{\citenamefont {Finco}\ \emph {et~al.}(2024)\citenamefont {Finco},
  \citenamefont {Li}, \citenamefont {Pohl}, \citenamefont {Reig~Escal{\'e}},
  \citenamefont {Maeder}, \citenamefont {Kaufmann},\ and\ \citenamefont
  {Grange}}]{finco2024monolithic}%
  \BibitemOpen
  \bibfield  {author} {\bibinfo {author} {\bibfnamefont {G.}~\bibnamefont
  {Finco}}, \bibinfo {author} {\bibfnamefont {G.}~\bibnamefont {Li}}, \bibinfo
  {author} {\bibfnamefont {D.}~\bibnamefont {Pohl}}, \bibinfo {author}
  {\bibfnamefont {M.}~\bibnamefont {Reig~Escal{\'e}}}, \bibinfo {author}
  {\bibfnamefont {A.}~\bibnamefont {Maeder}}, \bibinfo {author} {\bibfnamefont
  {F.}~\bibnamefont {Kaufmann}}, \ and\ \bibinfo {author} {\bibfnamefont
  {R.}~\bibnamefont {Grange}},\ }\bibfield  {title} {\enquote {\bibinfo {title}
  {Monolithic thin-film lithium niobate broadband spectrometer with one
  nanometre resolution},}\ }\href {\doibase
  https://doi.org/10.1038/s41467-024-46512-4} {\bibfield  {journal} {\bibinfo
  {journal} {Nat. Commun.}\ }\textbf {\bibinfo {volume} {15}},\ \bibinfo
  {pages} {2330} (\bibinfo {year} {2024})}\BibitemShut {NoStop}%
\bibitem [{\citenamefont {Zheng}\ \emph {et~al.}(2019)\citenamefont {Zheng},
  \citenamefont {Zou}, \citenamefont {Cai}, \citenamefont {Song}, \citenamefont
  {Chin}, \citenamefont {Liu}, \citenamefont {Lin}, \citenamefont {Kwong},\
  and\ \citenamefont {Liu}}]{zheng2019microring}%
  \BibitemOpen
  \bibfield  {author} {\bibinfo {author} {\bibfnamefont {S.~N.}\ \bibnamefont
  {Zheng}}, \bibinfo {author} {\bibfnamefont {J.}~\bibnamefont {Zou}}, \bibinfo
  {author} {\bibfnamefont {H.}~\bibnamefont {Cai}}, \bibinfo {author}
  {\bibfnamefont {J.}~\bibnamefont {Song}}, \bibinfo {author} {\bibfnamefont
  {L.}~\bibnamefont {Chin}}, \bibinfo {author} {\bibfnamefont {P.}~\bibnamefont
  {Liu}}, \bibinfo {author} {\bibfnamefont {Z.}~\bibnamefont {Lin}}, \bibinfo
  {author} {\bibfnamefont {D.}~\bibnamefont {Kwong}}, \ and\ \bibinfo {author}
  {\bibfnamefont {A.~Q.}\ \bibnamefont {Liu}},\ }\bibfield  {title} {\enquote
  {\bibinfo {title} {Microring resonator-assisted fourier transform
  spectrometer with enhanced resolution and large bandwidth in single chip
  solution},}\ }\href {\doibase https://doi.org/10.1038/s41467-019-10282-1}
  {\bibfield  {journal} {\bibinfo  {journal} {Nat. Commun.}\ }\textbf {\bibinfo
  {volume} {10}},\ \bibinfo {pages} {2349} (\bibinfo {year}
  {2019})}\BibitemShut {NoStop}%
\bibitem [{\citenamefont {Liu}\ \emph {et~al.}(2019)\citenamefont {Liu},
  \citenamefont {Li},\ and\ \citenamefont {Li}}]{liu2019electromechanical}%
  \BibitemOpen
  \bibfield  {author} {\bibinfo {author} {\bibfnamefont {Q.}~\bibnamefont
  {Liu}}, \bibinfo {author} {\bibfnamefont {H.}~\bibnamefont {Li}}, \ and\
  \bibinfo {author} {\bibfnamefont {M.}~\bibnamefont {Li}},\ }\bibfield
  {title} {\enquote {\bibinfo {title} {Electromechanical brillouin scattering
  in integrated optomechanical waveguides},}\ }\href {\doibase
  https://doi.org/10.1038/ncomms6402} {\bibfield  {journal} {\bibinfo
  {journal} {Optica}\ }\textbf {\bibinfo {volume} {6}},\ \bibinfo {pages} {778}
  (\bibinfo {year} {2019})}\BibitemShut {NoStop}%
\bibitem [{\citenamefont {Cheben}\ \emph {et~al.}(2007)\citenamefont {Cheben},
  \citenamefont {Schmid}, \citenamefont {Del{\^a}ge}, \citenamefont {Densmore},
  \citenamefont {Janz}, \citenamefont {Lamontagne}, \citenamefont {Lapointe},
  \citenamefont {Post}, \citenamefont {Waldron},\ and\ \citenamefont
  {Xu}}]{cheben2007high}%
  \BibitemOpen
  \bibfield  {author} {\bibinfo {author} {\bibfnamefont {P.}~\bibnamefont
  {Cheben}}, \bibinfo {author} {\bibfnamefont {J.}~\bibnamefont {Schmid}},
  \bibinfo {author} {\bibfnamefont {A.}~\bibnamefont {Del{\^a}ge}}, \bibinfo
  {author} {\bibfnamefont {A.}~\bibnamefont {Densmore}}, \bibinfo {author}
  {\bibfnamefont {S.}~\bibnamefont {Janz}}, \bibinfo {author} {\bibfnamefont
  {B.}~\bibnamefont {Lamontagne}}, \bibinfo {author} {\bibfnamefont
  {J.}~\bibnamefont {Lapointe}}, \bibinfo {author} {\bibfnamefont
  {E.}~\bibnamefont {Post}}, \bibinfo {author} {\bibfnamefont {P.}~\bibnamefont
  {Waldron}}, \ and\ \bibinfo {author} {\bibfnamefont {D.-X.}\ \bibnamefont
  {Xu}},\ }\bibfield  {title} {\enquote {\bibinfo {title} {A high-resolution
  silicon-on-insulator arrayed waveguide grating microspectrometer with
  sub-micrometer aperture waveguides},}\ }\href {\doibase
  https://doi.org/10.1364/OE.15.002299} {\bibfield  {journal} {\bibinfo
  {journal} {Opt. Express}\ }\textbf {\bibinfo {volume} {15}},\ \bibinfo
  {pages} {2299} (\bibinfo {year} {2007})}\BibitemShut {NoStop}%
\bibitem [{\citenamefont {Koshelev}\ \emph {et~al.}(2014)\citenamefont
  {Koshelev}, \citenamefont {Calafiore}, \citenamefont {Peroz}, \citenamefont
  {Dhuey}, \citenamefont {Cabrini}, \citenamefont {Sasorov}, \citenamefont
  {Goltsov},\ and\ \citenamefont {Yankov}}]{koshelev2014combination}%
  \BibitemOpen
  \bibfield  {author} {\bibinfo {author} {\bibfnamefont {A.}~\bibnamefont
  {Koshelev}}, \bibinfo {author} {\bibfnamefont {G.}~\bibnamefont {Calafiore}},
  \bibinfo {author} {\bibfnamefont {C.}~\bibnamefont {Peroz}}, \bibinfo
  {author} {\bibfnamefont {S.}~\bibnamefont {Dhuey}}, \bibinfo {author}
  {\bibfnamefont {S.}~\bibnamefont {Cabrini}}, \bibinfo {author} {\bibfnamefont
  {P.}~\bibnamefont {Sasorov}}, \bibinfo {author} {\bibfnamefont
  {A.}~\bibnamefont {Goltsov}}, \ and\ \bibinfo {author} {\bibfnamefont
  {V.}~\bibnamefont {Yankov}},\ }\bibfield  {title} {\enquote {\bibinfo {title}
  {Combination of a spectrometer-on-chip and an array of young’s
  interferometers for laser spectrum monitoring},}\ }\href {\doibase
  https://doi.org/10.1364/OL.39.005645} {\bibfield  {journal} {\bibinfo
  {journal} {Opt. Lett.}\ }\textbf {\bibinfo {volume} {39}},\ \bibinfo {pages}
  {5645} (\bibinfo {year} {2014})}\BibitemShut {NoStop}%
\bibitem [{\citenamefont {Zou}\ \emph {et~al.}(2016)\citenamefont {Zou},
  \citenamefont {Lang}, \citenamefont {Le},\ and\ \citenamefont
  {He}}]{zou2016ultracompact}%
  \BibitemOpen
  \bibfield  {author} {\bibinfo {author} {\bibfnamefont {J.}~\bibnamefont
  {Zou}}, \bibinfo {author} {\bibfnamefont {T.}~\bibnamefont {Lang}}, \bibinfo
  {author} {\bibfnamefont {Z.}~\bibnamefont {Le}}, \ and\ \bibinfo {author}
  {\bibfnamefont {J.-J.}\ \bibnamefont {He}},\ }\bibfield  {title} {\enquote
  {\bibinfo {title} {Ultracompact silicon-on-insulator-based reflective arrayed
  waveguide gratings for spectroscopic applications},}\ }\href {\doibase
  https://doi.org/10.1364/AO.55.003531} {\bibfield  {journal} {\bibinfo
  {journal} {Appl. Opt.}\ }\textbf {\bibinfo {volume} {55}},\ \bibinfo {pages}
  {3531} (\bibinfo {year} {2016})}\BibitemShut {NoStop}%
\bibitem [{\citenamefont {Nedeljkovic}\ \emph {et~al.}(2015)\citenamefont
  {Nedeljkovic}, \citenamefont {Velasco}, \citenamefont {Khokhar},
  \citenamefont {Delage}, \citenamefont {Cheben},\ and\ \citenamefont
  {Mashanovich}}]{nedeljkovic2015mid}%
  \BibitemOpen
  \bibfield  {author} {\bibinfo {author} {\bibfnamefont {M.}~\bibnamefont
  {Nedeljkovic}}, \bibinfo {author} {\bibfnamefont {A.~V.}\ \bibnamefont
  {Velasco}}, \bibinfo {author} {\bibfnamefont {A.~Z.}\ \bibnamefont
  {Khokhar}}, \bibinfo {author} {\bibfnamefont {A.}~\bibnamefont {Delage}},
  \bibinfo {author} {\bibfnamefont {P.}~\bibnamefont {Cheben}}, \ and\ \bibinfo
  {author} {\bibfnamefont {G.~Z.}\ \bibnamefont {Mashanovich}},\ }\bibfield
  {title} {\enquote {\bibinfo {title} {Mid-infrared silicon-on-insulator
  fourier-transform spectrometer chip},}\ }\href {\doibase
  10.1109/LPT.2015.2496729} {\bibfield  {journal} {\bibinfo  {journal} {IEEE
  Photon. Technol. Lett.}\ }\textbf {\bibinfo {volume} {28}},\ \bibinfo {pages}
  {528} (\bibinfo {year} {2015})}\BibitemShut {NoStop}%
\bibitem [{\citenamefont {Nie}\ \emph {et~al.}(2017)\citenamefont {Nie},
  \citenamefont {Ryckeboer}, \citenamefont {Roelkens},\ and\ \citenamefont
  {Baets}}]{nie2017cmos}%
  \BibitemOpen
  \bibfield  {author} {\bibinfo {author} {\bibfnamefont {X.}~\bibnamefont
  {Nie}}, \bibinfo {author} {\bibfnamefont {E.}~\bibnamefont {Ryckeboer}},
  \bibinfo {author} {\bibfnamefont {G.}~\bibnamefont {Roelkens}}, \ and\
  \bibinfo {author} {\bibfnamefont {R.}~\bibnamefont {Baets}},\ }\bibfield
  {title} {\enquote {\bibinfo {title} {Cmos-compatible broadband co-propagative
  stationary fourier transform spectrometer integrated on a silicon nitride
  photonics platform},}\ }\href {\doibase https://doi.org/10.1364/OE.25.00A409}
  {\bibfield  {journal} {\bibinfo  {journal} {Opt. Express}\ }\textbf {\bibinfo
  {volume} {25}},\ \bibinfo {pages} {A409} (\bibinfo {year}
  {2017})}\BibitemShut {NoStop}%
\bibitem [{\citenamefont {Wang}\ \emph {et~al.}(2019)\citenamefont {Wang},
  \citenamefont {Yi}, \citenamefont {Chen}, \citenamefont {Zhou}, \citenamefont
  {Luk}, \citenamefont {James}, \citenamefont {Nogan}, \citenamefont {Ross},
  \citenamefont {Joe}, \citenamefont {Shahsafi} \emph
  {et~al.}}]{wang2019single}%
  \BibitemOpen
  \bibfield  {author} {\bibinfo {author} {\bibfnamefont {Z.}~\bibnamefont
  {Wang}}, \bibinfo {author} {\bibfnamefont {S.}~\bibnamefont {Yi}}, \bibinfo
  {author} {\bibfnamefont {A.}~\bibnamefont {Chen}}, \bibinfo {author}
  {\bibfnamefont {M.}~\bibnamefont {Zhou}}, \bibinfo {author} {\bibfnamefont
  {T.~S.}\ \bibnamefont {Luk}}, \bibinfo {author} {\bibfnamefont
  {A.}~\bibnamefont {James}}, \bibinfo {author} {\bibfnamefont
  {J.}~\bibnamefont {Nogan}}, \bibinfo {author} {\bibfnamefont
  {W.}~\bibnamefont {Ross}}, \bibinfo {author} {\bibfnamefont {G.}~\bibnamefont
  {Joe}}, \bibinfo {author} {\bibfnamefont {A.}~\bibnamefont {Shahsafi}},
  \emph {et~al.},\ }\bibfield  {title} {\enquote {\bibinfo {title} {Single-shot
  on-chip spectral sensors based on photonic crystal slabs},}\ }\href {\doibase
  https://doi.org/10.1038/s41467-019-08994-5} {\bibfield  {journal} {\bibinfo
  {journal} {Nat. Commun.}\ }\textbf {\bibinfo {volume} {10}},\ \bibinfo
  {pages} {1020} (\bibinfo {year} {2019})}\BibitemShut {NoStop}%
\bibitem [{\citenamefont {Redding}\ \emph {et~al.}(2013)\citenamefont
  {Redding}, \citenamefont {Liew}, \citenamefont {Sarma},\ and\ \citenamefont
  {Cao}}]{redding2013compact}%
  \BibitemOpen
  \bibfield  {author} {\bibinfo {author} {\bibfnamefont {B.}~\bibnamefont
  {Redding}}, \bibinfo {author} {\bibfnamefont {S.~F.}\ \bibnamefont {Liew}},
  \bibinfo {author} {\bibfnamefont {R.}~\bibnamefont {Sarma}}, \ and\ \bibinfo
  {author} {\bibfnamefont {H.}~\bibnamefont {Cao}},\ }\bibfield  {title}
  {\enquote {\bibinfo {title} {Compact spectrometer based on a disordered
  photonic chip},}\ }\href {\doibase https://doi.org/10.1038/nphoton.2013.190}
  {\bibfield  {journal} {\bibinfo  {journal} {Nat. Photon.}\ }\textbf {\bibinfo
  {volume} {7}},\ \bibinfo {pages} {746} (\bibinfo {year} {2013})}\BibitemShut
  {NoStop}%
\bibitem [{\citenamefont {Yang}\ \emph {et~al.}(2019)\citenamefont {Yang},
  \citenamefont {Albrow-Owen}, \citenamefont {Cui}, \citenamefont
  {Alexander-Webber}, \citenamefont {Gu}, \citenamefont {Wang}, \citenamefont
  {Wu}, \citenamefont {Zhuge}, \citenamefont {Williams}, \citenamefont {Wang}
  \emph {et~al.}}]{yang2019single}%
  \BibitemOpen
  \bibfield  {author} {\bibinfo {author} {\bibfnamefont {Z.}~\bibnamefont
  {Yang}}, \bibinfo {author} {\bibfnamefont {T.}~\bibnamefont {Albrow-Owen}},
  \bibinfo {author} {\bibfnamefont {H.}~\bibnamefont {Cui}}, \bibinfo {author}
  {\bibfnamefont {J.}~\bibnamefont {Alexander-Webber}}, \bibinfo {author}
  {\bibfnamefont {F.}~\bibnamefont {Gu}}, \bibinfo {author} {\bibfnamefont
  {X.}~\bibnamefont {Wang}}, \bibinfo {author} {\bibfnamefont {T.-C.}\
  \bibnamefont {Wu}}, \bibinfo {author} {\bibfnamefont {M.}~\bibnamefont
  {Zhuge}}, \bibinfo {author} {\bibfnamefont {C.}~\bibnamefont {Williams}},
  \bibinfo {author} {\bibfnamefont {P.}~\bibnamefont {Wang}},  \emph {et~al.},\
  }\bibfield  {title} {\enquote {\bibinfo {title} {Single-nanowire
  spectrometers},}\ }\href {\doibase 10.1126/science.aax8814} {\bibfield
  {journal} {\bibinfo  {journal} {Science}\ }\textbf {\bibinfo {volume}
  {365}},\ \bibinfo {pages} {1017} (\bibinfo {year} {2019})}\BibitemShut
  {NoStop}%
\bibitem [{\citenamefont {He}\ \emph {et~al.}(2024)\citenamefont {He},
  \citenamefont {Li}, \citenamefont {Yu}, \citenamefont {Zhou}, \citenamefont
  {Ke}, \citenamefont {Yip},\ and\ \citenamefont {Zhao}}]{he2024microsized}%
  \BibitemOpen
  \bibfield  {author} {\bibinfo {author} {\bibfnamefont {X.}~\bibnamefont
  {He}}, \bibinfo {author} {\bibfnamefont {Y.}~\bibnamefont {Li}}, \bibinfo
  {author} {\bibfnamefont {H.}~\bibnamefont {Yu}}, \bibinfo {author}
  {\bibfnamefont {G.}~\bibnamefont {Zhou}}, \bibinfo {author} {\bibfnamefont
  {L.}~\bibnamefont {Ke}}, \bibinfo {author} {\bibfnamefont {H.-L.}\
  \bibnamefont {Yip}}, \ and\ \bibinfo {author} {\bibfnamefont
  {N.}~\bibnamefont {Zhao}},\ }\bibfield  {title} {\enquote {\bibinfo {title}
  {A microsized optical spectrometer based on an organic photodetector with an
  electrically tunable spectral response},}\ }\href {\doibase
  https://doi.org/10.1038/s41928-024-01199-9} {\bibfield  {journal} {\bibinfo
  {journal} {Nat. Electron.}\ ,\ \bibinfo {pages} {1}} (\bibinfo {year}
  {2024})}\BibitemShut {NoStop}%
\bibitem [{\citenamefont {Li}\ and\ \citenamefont
  {Fainman}(2021)}]{li2021chip}%
  \BibitemOpen
  \bibfield  {author} {\bibinfo {author} {\bibfnamefont {A.}~\bibnamefont
  {Li}}\ and\ \bibinfo {author} {\bibfnamefont {Y.}~\bibnamefont {Fainman}},\
  }\bibfield  {title} {\enquote {\bibinfo {title} {On-chip spectrometers using
  stratified waveguide filters},}\ }\href {\doibase
  https://doi.org/10.1038/s41467-021-23001-6} {\bibfield  {journal} {\bibinfo
  {journal} {Nat. Commun.}\ }\textbf {\bibinfo {volume} {12}},\ \bibinfo
  {pages} {2704} (\bibinfo {year} {2021})}\BibitemShut {NoStop}%
\bibitem [{\citenamefont {Sipe}\ and\ \citenamefont
  {Steel}(2016)}]{sipe2016hamiltonian}%
  \BibitemOpen
  \bibfield  {author} {\bibinfo {author} {\bibfnamefont {J.}~\bibnamefont
  {Sipe}}\ and\ \bibinfo {author} {\bibfnamefont {M.}~\bibnamefont {Steel}},\
  }\bibfield  {title} {\enquote {\bibinfo {title} {A hamiltonian treatment of
  stimulated brillouin scattering in nanoscale integrated waveguides},}\ }\href
  {\doibase 10.1088/1367-2630/18/4/045004} {\bibfield  {journal} {\bibinfo
  {journal} {New J. Phys.}\ }\textbf {\bibinfo {volume} {18}},\ \bibinfo
  {pages} {045004} (\bibinfo {year} {2016})}\BibitemShut {NoStop}%
\bibitem [{\citenamefont {Bajak}\ \emph {et~al.}(1981)\citenamefont {Bajak},
  \citenamefont {McNab}, \citenamefont {Richter},\ and\ \citenamefont
  {Wilkinson}}]{1981attenuation}%
  \BibitemOpen
  \bibfield  {author} {\bibinfo {author} {\bibfnamefont {I.}~\bibnamefont
  {Bajak}}, \bibinfo {author} {\bibfnamefont {A.}~\bibnamefont {McNab}},
  \bibinfo {author} {\bibfnamefont {J.}~\bibnamefont {Richter}}, \ and\
  \bibinfo {author} {\bibfnamefont {C.}~\bibnamefont {Wilkinson}},\ }\bibfield
  {title} {\enquote {\bibinfo {title} {Attenuation of acoustic waves in lithium
  niobate},}\ }\href {\doibase https://doi.org/10.1121/1.385588} {\bibfield
  {journal} {\bibinfo  {journal} {J. Acoust. Soc. Am.}\ }\textbf {\bibinfo
  {volume} {69}},\ \bibinfo {pages} {689} (\bibinfo {year} {1981})}\BibitemShut
  {NoStop}%
\bibitem [{\citenamefont {Chen}\ \emph {et~al.}(2024)\citenamefont {Chen},
  \citenamefont {Briggs}, \citenamefont {Cui}, \citenamefont {Zhang},
  \citenamefont {Shah},\ and\ \citenamefont {Fan}}]{linran2024adapted}%
  \BibitemOpen
  \bibfield  {author} {\bibinfo {author} {\bibfnamefont {P.-K.}\ \bibnamefont
  {Chen}}, \bibinfo {author} {\bibfnamefont {I.}~\bibnamefont {Briggs}},
  \bibinfo {author} {\bibfnamefont {C.}~\bibnamefont {Cui}}, \bibinfo {author}
  {\bibfnamefont {L.}~\bibnamefont {Zhang}}, \bibinfo {author} {\bibfnamefont
  {M.}~\bibnamefont {Shah}}, \ and\ \bibinfo {author} {\bibfnamefont
  {L.}~\bibnamefont {Fan}},\ }\bibfield  {title} {\enquote {\bibinfo {title}
  {Adapted poling to break the nonlinear efficiency limit in nanophotonic
  lithium niobate waveguides},}\ }\href {\doibase
  https://doi.org/10.1038/s41565-023-01525-w} {\bibfield  {journal} {\bibinfo
  {journal} {Nat. Nano.}\ }\textbf {\bibinfo {volume} {19}},\ \bibinfo {pages}
  {44} (\bibinfo {year} {2024})}\BibitemShut {NoStop}%
\bibitem [{\citenamefont {Xu}\ \emph {et~al.}(2022)\citenamefont {Xu},
  \citenamefont {Wang}, \citenamefont {Sun},\ and\ \citenamefont
  {Zou}}]{xu2022hybrid}%
  \BibitemOpen
  \bibfield  {author} {\bibinfo {author} {\bibfnamefont {X.-B.}\ \bibnamefont
  {Xu}}, \bibinfo {author} {\bibfnamefont {W.-T.}\ \bibnamefont {Wang}},
  \bibinfo {author} {\bibfnamefont {L.-Y.}\ \bibnamefont {Sun}}, \ and\
  \bibinfo {author} {\bibfnamefont {C.-L.}\ \bibnamefont {Zou}},\ }\bibfield
  {title} {\enquote {\bibinfo {title} {Hybrid superconducting photonic-phononic
  chip for quantum information processing},}\ }\href {\doibase
  https://doi.org/10.1016/j.chip.2022.100016} {\bibfield  {journal} {\bibinfo
  {journal} {Chip}\ }\textbf {\bibinfo {volume} {1}},\ \bibinfo {pages}
  {100016} (\bibinfo {year} {2022})}\BibitemShut {NoStop}%
\bibitem [{\citenamefont {Chen}\ \emph {et~al.}(2022)\citenamefont {Chen},
  \citenamefont {Liu}, \citenamefont {Li}, \citenamefont {Song}, \citenamefont
  {Wang}, \citenamefont {Wang}, \citenamefont {Wang},\ and\ \citenamefont
  {Zhu}}]{chen2022scandium}%
  \BibitemOpen
  \bibfield  {author} {\bibinfo {author} {\bibfnamefont {L.}~\bibnamefont
  {Chen}}, \bibinfo {author} {\bibfnamefont {C.}~\bibnamefont {Liu}}, \bibinfo
  {author} {\bibfnamefont {M.}~\bibnamefont {Li}}, \bibinfo {author}
  {\bibfnamefont {W.}~\bibnamefont {Song}}, \bibinfo {author} {\bibfnamefont
  {W.}~\bibnamefont {Wang}}, \bibinfo {author} {\bibfnamefont {Z.}~\bibnamefont
  {Wang}}, \bibinfo {author} {\bibfnamefont {N.}~\bibnamefont {Wang}}, \ and\
  \bibinfo {author} {\bibfnamefont {Y.}~\bibnamefont {Zhu}},\ }\bibfield
  {title} {\enquote {\bibinfo {title} {Scandium-doped aluminum nitride for
  acoustic wave resonators, filters, and ferroelectric memory applications},}\
  }\href {\doibase https://doi.org/10.1021/acsaelm.2c01409} {\bibfield
  {journal} {\bibinfo  {journal} {ACS Appl. Electron. Mater.}\ }\textbf
  {\bibinfo {volume} {5}},\ \bibinfo {pages} {612} (\bibinfo {year}
  {2022})}\BibitemShut {NoStop}%
\bibitem [{\citenamefont {Bartasyte}\ \emph {et~al.}(2017)\citenamefont
  {Bartasyte}, \citenamefont {Margueron}, \citenamefont {Baron}, \citenamefont
  {Oliveri},\ and\ \citenamefont {Boulet}}]{bartasyte2017toward}%
  \BibitemOpen
  \bibfield  {author} {\bibinfo {author} {\bibfnamefont {A.}~\bibnamefont
  {Bartasyte}}, \bibinfo {author} {\bibfnamefont {S.}~\bibnamefont
  {Margueron}}, \bibinfo {author} {\bibfnamefont {T.}~\bibnamefont {Baron}},
  \bibinfo {author} {\bibfnamefont {S.}~\bibnamefont {Oliveri}}, \ and\
  \bibinfo {author} {\bibfnamefont {P.}~\bibnamefont {Boulet}},\ }\bibfield
  {title} {\enquote {\bibinfo {title} {Toward high-quality epitaxial
  {L}i{N}b{O}3 and {L}i{T}a{O}3 thin films for acoustic and optical
  applications},}\ }\href {\doibase https://doi.org/10.1002/admi.201600998}
  {\bibfield  {journal} {\bibinfo  {journal} {Adv. Mater. Interfaces}\ }\textbf
  {\bibinfo {volume} {4}},\ \bibinfo {pages} {1600998} (\bibinfo {year}
  {2017})}\BibitemShut {NoStop}%
\end{thebibliography}
\end{document}